\documentclass[%
aip,
 amsmath,amssymb,
preprint,%
]{revtex4-1}
\usepackage{accents}
\usepackage{xcolor}
\usepackage{subfig}
\usepackage{graphicx}
\usepackage{dcolumn}
\usepackage{amsmath,amsfonts,amsthm,bm}
\usepackage{appendix}
\usepackage{color}
\usepackage{float}
\usepackage[colorinlistoftodos]{todonotes}

\begin{document}


   \title{Ion temperature effects on plasma flow in the magnetic mirror configuration}

\author{A. Sabo}\altaffiliation{Corresponding Author}%
 \email{ans543@usask.ca}
\affiliation{University of Saskatchewan, Saskatchewan, Saskatoon SK S7N 5E2, Canada 
}

\author{A. I. Smolyakov}%
\affiliation{%
University of Saskatchewan, Saskatchewan, Saskatoon SK S7N 5E2, Canada 
}%

\author{P. Yushmanov}%
\affiliation{%
 TAE Technologies, 19631 Pauling, Foothill Ranch, CA, 92610, United States 
}%

\author{S. Putvinski}%
\affiliation{%
 TAE Technologies, 19631 Pauling, Foothill Ranch, CA, 92610, United States 
}%

\begin{abstract}
Effects of finite ion temperature on plasma flow in the converging-diverging magnetic field, the magnetic mirror, or equivalently,  magnetic nozzle configuration, are studied using a quasineutral paraxial two-fluid MHD model with isothermal electrons and  warm magnetized ions.   The ion acceleration was studied with an emphasis on the role of the singularity at the sonic point transition. It is shown that the regularity of the sonic point defines a global solution describing plasma acceleration from subsonic to supersonic velocity. Stationary accelerating solutions  were obtained and compared with the time dependent dynamics, confirming that the solutions of the time-dependent equations converge to the stationary solutions and therefore are stable. The effects of the ion pressure anisotropy  were analyzed using the Chew-Golberger-Low model and its generalization. It is shown that the mirror force (manifested by the perpendicular ion pressure)  enhances plasma acceleration. The role of  ionization and charge exchange on plasma flow acceleration have been investigated.
\end{abstract}

\keywords{Plasma acceleration, magnetic nozzle, anisotropic pressure, CGL,
mirror machine}
\maketitle


\affiliation{ University of Saskatchewan, Saskatchewan, Saskatoon SK S7N 5E2, Canada
}

\affiliation{University of Saskatchewan, Saskatchewan, Saskatoon SK S7N 5E2, Canada
}%


%

\section{Introduction}

Plasma flow in the magnetic mirror configuration (magnetic
nozzle) plays an important role in a number of devices, such as magnetic
mirrors used in fusion research and devices for electric propulsion in space.
In plasma propulsion, the magnetic nozzle configuration is employed to  convert the plasma thermal energy  into the ion kinetic
energy, thus generating thrust \cite{KaganovichPoP2020,LongmierPSST2011}. In open mirror fusion devices,  the expanding magnetic field of  the divertor (expander region) is used  to spread the energy over the larger area to reduce the wall heat loads  
\cite{onofri2017magnetohydrodynamic}. Plasma flow in the edge region of the divertor tokamak also experiences acceleration to supersonic velocities due to the combined effects of plasma pressure and inhomogeneous magnetic field  \cite{BufferandPPCF2014,KirkPPCF2003,GhendrihPPCF2011}. 
Various aspects of plasma flow and acceleration in the magnetic mirror configurations  have been studied. General framework of plasma flow and acceleration in the MHD approximation was formulated in Ref. \onlinecite{morozov1980steady}, and subsequently used to study plasma detachment \cite{HooperAIAA1993},   acceleration mechanisms and propulsion efficiency  \cite{AhedoPoP2010,MerinoPoP2016}. The role of the magnetic nozzle in the conversion of plasma thermal energy to supersonic flow was  demonstrated experimentally \cite{InutakePST2004,InutakePPCF2007}.  The MHD models were used to study the supersonic acceleration in the scrape-off-layer (SOL) of tokamak edge in Refs. \onlinecite{GoswamiPoP2014,TogoCPP2018,TogoNF2019}.
Ion pressure anisotropy  and related mirror force effects on the plasma flow in the SOL were considered and the predictions of the isotropic and anisotropic pressure models were compared for specific conditions of the advanced divertors in fusion systems in Refs. \onlinecite{TogoCPP2018,TogoNF2019}. The emphasis of this paper is on the role of the sonic point transition in the formation of the accelerating  potential in  plasma with anisotropic ion pressure and including atomic processes such as ionization and charge exchange.  
 
 It is well known that in the quasineutral approximation the ion inertia and finite temperature will  result in the appearance of the sonic point singularity at the point where the local ion velocity is equal to the ion sound speed $c_s$.
 It has recently been emphasized \cite{SmolyakovPoP2021} that  the conditions at the sonic point in the nozzle region where the magnetic field has the maximum are critical for the existence of the smooth accelerating solution such that the  resulting  plasma flow is uniquely defined in the whole converging-diverging region, i.e. in the whole range from sub-sonic,  $V<c_s$, to super-sonic,  $V>c_s$, velocities. 
Such smooth accelerating solution is formed by the ambipolar potential supported by the electron pressure gradient. 
Here we study how the addition of the ion pressure, its anisotropy  in  particular, the effects of the mirror force,  and some dissipative processes modify the conditions for the formation of a smooth accelerating potential. 
These results are  of interest for fusion devices application \cite{TogoNF2019,onofri2017magnetohydrodynamic} as well as for propulsion devices where a large ion temperature is expected \cite{LongmierPSST2011}.
 
 The basic model equations are presented in Section II.  The Section III discusses general features of the  sonic point singularity and ion acceleration with finite ion pressure. Plasma acceleration for cold ions is reviewed in Section IV.  The Section V presents the results of the solution of the stationary and dynamic (time-dependent) equations for the base case with anisotropic pressure.  The Section VI analyzes modifications due to the ionization and charge-exchange collisions. The summary and discussions are presented in Section VII.

\section{Basic model}

In our model, electrons are considered in the massless  isothermal $T_{e}=const$ approximation,  
 \begin{equation}
0=en\frac{\partial \phi }{\partial z}-T_{e}\frac{\partial n}{\partial z}.
\label{potential-gradient}
\end{equation}%
The ions are described by the two-pressure  Chew-Golberger-Low (CGL) model \cite{chew1956boltzmann} in the form   
 \begin{equation}
\frac{d}{dt}\left( \frac{p_{\parallel }B^{2}}{n^{3}}\right) =S_\parallel,
\label{p2-para}
\end{equation}%
\begin{equation}
\frac{d}{dt}\left( \frac{p_{\perp }}{nB}\right)   =S_\perp,
\label{p2-perp}
\end{equation}
where $p_\perp$ and $p_\Vert$
are the perpendicular and parallel ion pressure, respectively. The standard  CGL model is generalized here to include the sink terms $S_\perp$, $S_\Vert$ due to dissipation related to ionization and charge-exchange.
 We consider a problem in the paraxial approximation so that  the total fluid time derivative for ions is $d/dt=\partial /\partial t+V_{\parallel } {\nabla }_\parallel
$, where $V_{\parallel }=\mathbf{V }\cdot \mathbf{B}/B$ is the ion parallel velocity, ${\nabla }_\parallel =\mathbf{b}\cdot \nabla= \mathbf{B}/B\cdot \nabla$, and $\mathbf{b}=\mathbf{B}/B$ represents a unit vector in the direction of the magnetic field. 
 
In the paraxial approximation, the 
ion continuity and momentum equations take the form
\begin{equation}
\frac{\partial n}{\partial t} +  B \frac{\partial}{\partial z} \frac{ n V_\parallel }{B}=S_n,
\end{equation}
  \begin{equation}
m_{i}n\left( \frac{\partial V_{\parallel }}{\partial t}+V_{\parallel }\frac{%
\partial V_{\parallel }}{\partial z}\right) =  -en\frac{\partial \phi }{%
\partial z}-\nabla p- {\mathbf {b}}\cdot \nabla {\boldsymbol  {\pi}}+S_V.
\end{equation}
The effects of anisotropic pressure are included in the momentum equation with  anisotropic pressure tensor ${\boldsymbol  {\pi}}=(p_\Vert-p_\perp)({\mathbf b} {\mathbf b} -{\mathbf I}/3)$, $p=(2p_\perp+p_\Vert)/3$,. and where $ S_n, S_V$ describe the ionization source/sink effects. 
 
 In the context of the plasma flow and acceleration along the open magnetic field lines, the anisotropic MHD equations  were considered in Refs.  \onlinecite{zawaideh1986generalized,GuoPoP2014}. Similar equations also follow from the  drift-kinetic \cite{SmolyakovPoP2010} or gyro-fluid  equations \cite{SnyderPoP1997,RobertsonPoP2016}.  In  simulations of plasma flow in the SOL tokamak regions, the fluid models with additional source/sink terms in the density, momentum, and energy equations as well as model approximations for the heat fluxes and relaxation terms were used \cite{TogoNF2019}. In this paper, our emphasis is on the ion pressure effects, thus we use the standard adiabatic CGL equations  \cite{chew1956boltzmann} modified  with the model sink terms $S_\parallel$ and $S_\perp$ to  model the effects of ionization and charge-exchange on the ion pressure evolution,  cf.   Eqs. (\ref{p2-para}) and (\ref{p2-perp}). 
 
   In general, ionization contributes to the density, ion momentum and pressure evolution, while the charge exchange affects the ion momentum and pressure evolution. In the context of fusion and electric propulsion, the ionization by electron impact is most relevant.  The ionization  coefficients depend on the neutral species, neutral density, electron temperature and electron density, while  charge-exchange coefficients depend on the ions/neutral density and their energy. Thus the ionization and charge-exchange  effects may have complex profile dependencies depending on the particular situation and plasma parameters. In this study we are interested in the main parametric trends due to the ionization and charge exchange. Thus, we use a simplified model \cite{HelanderPoP1994,NgPoP2007}  where the effects of ionization and charge exchange are represented by two constant coefficients $\nu_1$ and $\nu_{2}$. The $\nu_1$ coefficient corresponds to the ionization, while the $\nu_2$ coefficient in the ion momentum equation describes the total effects of ionization and charge-exchange. In this study we neglect the heat flux effects in the energy (pressure evolution) as well as any possible ionization heating terms \cite{LehnertNF1971} and only the pressure "decay" terms are  included with the same coefficient $\nu_2$ for $p_\Vert$ and $p_\perp$. In the expanded form, the full system of equations including ionization and charge-exchange effects is then written in the form similarly to Ref. \onlinecite{onofri2017magnetohydrodynamic}: 
  
\begin{equation}
\frac{\partial n}{\partial t} = n V_\parallel \frac{\partial \ln B}{%
\partial z} - V_\parallel \frac{\partial n}{\partial z} - n \frac{\partial
V_\parallel}{\partial z} + \nu_1 n,
\label{n-time-dependent-with-charge-exchange}
\end{equation}
\begin{equation}
\begin{split}
m_{i}n\left( \frac{\partial V_{\parallel }}{\partial t}+V_{\parallel }\frac{%
\partial V_{\parallel }}{\partial z}\right) = & -en\frac{\partial \phi }{%
\partial z}-\frac{\partial p_{\parallel }}{\partial z}+(p_{\parallel}-p_{\perp })\\
& \times\frac{\partial \ln B}{\partial z}-\nu _{2}m_{i}nV_{\Vert },
\end{split}
\label{momentum-eq-with-charge-exchange}
\end{equation}
\begin{equation}
\frac{\partial p_\parallel}{\partial t} = p_\parallel V_\parallel \frac{%
\partial \ln B}{\partial z} - V_\parallel \frac{\partial p_\parallel}{%
\partial z} - 3 p_\parallel \frac{\partial V_\parallel}{\partial z} -
\nu_2 p_\parallel,
\label{p-parallel-with-charge-exchange}
\end{equation}
\begin{equation}
\frac{\partial p_\perp}{\partial t} = 2 p_\perp V_\parallel \frac{\partial
\ln B}{\partial z} - V_\parallel \frac{\partial p_\perp}{\partial z} - p_\perp \frac{\partial V_\parallel}{\partial z} -\nu_2 p_\perp.
\label{p-perpendicular-with-charge-exchange}
\end{equation}

 We have to note that here we do not consider the plasma source region but focus on {\it plasma acceleration} in the magnetic mirror region,  $0<z<L$.  Thus,  effects of ionization and charge exchange,  represented by the $\nu_1$ and $\nu_2$ coefficients in the continuity, momentum, and pressure equations, are the  model approximations for the  ionization and charge-exchange  effects  in the magnetic mirror,  while  the plasma source is assumed for $z<0$.

Before we proceed with the general case, we consider general conditions for the existence of accelerating solution and describe  plasma acceleration for cold and warm ions in the absence of ionization and charge-exchange effects.

\section{The sonic point singularity and plasma acceleration}

In this section we consider a general condition for the existence of the global stationary accelerating solution in the absence of ionization and charge-exchange but taking into account anisotropic ion pressure. Setting $\nu_{1}=\nu_2=0$, for the stationary case one can obtain from     (\ref{potential-gradient} - \ref{p-perpendicular-with-charge-exchange}) the following equation for the ion velocity    \begin{equation}
\left( M^{2}-1-\frac{3p_{\parallel }}{nT_{e}}\right) \frac{\partial M}{%
\partial z}=-\left( 1+\frac{p_{\perp }}{nT_{e}}\right) M\frac{\partial \ln B%
}{\partial z}.  \label{gradient-of-M}
\end{equation}%
This equation has to be solved simultaneously with equations  (\ref{p-parallel-with-charge-exchange}) and (\ref{p-perpendicular-with-charge-exchange}) for $p_\perp$ and $p_\parallel$.
In equation  (\ref{gradient-of-M}), $M=V_{\parallel }/c_{s}$ is  the plasma velocity $V_{\parallel }$ normalized to the speed of sound with respect to the electron velocity  $c_{s}=\sqrt{(T_e/m_i)}$. Such normalization is  convenient because $T_e$ is constant and uniform. It is important to note that for warm  anisotropic ions, the actual sound velocity includes the parallel ion temperature, $v_{cs}=\sqrt{(T_e+3T_{i\Vert})/m_i}$. This is reflected in the modification of the sonic point   singularity  in equation (\ref{gradient-of-M}), which occurs at a point where the ion velocity is equal to the local ion sound velocity,  $M=M_{cr}=\sqrt{1+3p_{\parallel }/nT_{e}}$. 
 We note that the expression  $v_{cs}=\sqrt{(T_e+T_{i\Vert})/m_i}$ used for sound speed in some papers \cite{TogoNME2019,TogoNF2019} is incorrect and thus misrepresents the location of the transition point.  

Equation (\ref{gradient-of-M}) describes two mechanisms of the ion  acceleration: the electric field and the mirror force due to the perpendicular ion pressure. The electric field  is supported  by the electron pressure gradient in  Eq. (\ref{potential-gradient}), so this mechanism  comes from the electron thermal energy. The approximation of isothermal electrons used here  assumes an infinite source of electron energy. In practice, electrons can experience cooling along the flow \cite{LittlePRL2016,MerinoPSST2020} and thus their temperature is not constant. Semi-empirically, this  can be described by a general polytropic equation of state for electrons, $p_e\sim n^\gamma$. The ion acceleration due to the electron pressure with   $\gamma \neq 1$ was considered previously \cite{SmolyakovPoP2021} (not included here for simplicity).

An additional contribution to the ion acceleration comes from the ion perpendicular  pressure (energy),  manifested by the $p_\perp$ term on the right hand side of  (\ref{gradient-of-M}), and is the result of the mirror force. As it will be discussed below,  the effect of the ion parallel pressure  modifies the location of the sonic point in Eq. \ref{gradient-of-M} but the impact on the ion velocity is not significant due to the fast decrease of the ion pressure related to the decrease of  plasma density along the nozzle  and strong density dependence $\simeq n^3$ in the CGL equation for $p_\Vert$.

 In general,  acceleration in the magnetic mirror configuration is  similar to the Laval nozzle acceleration. The ion acceleration by the electric field is supported by the density drop. In subsonic region, for $M < M_{cr}$,  the ion inertia can be neglected and the acceleration occurs kinematically due to the effective area cross-section decrease as a result of the magnetic field rise,  $\partial \ln B/\partial z>0$. Effectively,  in this regime for $M\ll 1$, the density gradient is small and the acceleration simply follows from the flow conservation: $V^{-1}\partial V/\partial z \simeq  B^{-1} \partial B / \partial z$, under constant density.

Equation (\ref{gradient-of-M})
exhibits the singularity at $M=M_{cr}$, the point where the ion flow resonates with the ion sound mode. A global smooth accelerating solution  can  be obtained by regularizing the sonic point at the point $z=z_{m}$  where $\partial \ln B/\partial z=0$, which therefore requires the condition  
\begin{equation}
\left| M^{2}-1-\frac{3p_{\parallel }}{nT_{e}}\right\vert _{z=z_m} =0.
\label{M-sonic-point}
\end{equation}%

Expanding the
left and right sides of (\ref{gradient-of-M}) near the singular point and using equation (\ref{M-sonic-point}) one obtains the expression for
the $\partial M/\partial z$ derivative near $z=z_{m}:$
\begin{eqnarray}
\left( \frac{\partial M}{\partial z}\right) ^{2} &=&-\left. \frac{\left(
1+p_{\bot }/nT_{e}\right) \left( 1+3p_{\Vert }/nT_{e}\right) }{2\left(
1+6p_{\Vert }/nT_{e}\right) }\frac{\partial ^{2}\ln B}{\partial z^{2}}%
\right\vert _{z=z_{m}}  \notag \\
&=&-\left. \frac{\left( 1+T_{\bot }/T_{e}\right) \left( 1+3T_{\Vert
}/T_{e}\right) }{2\left( 1+6T_{\Vert }/T_{e}\right) }\frac{\partial ^{2}\ln B%
}{\partial z^{2}}\right\vert _{z=z_{m}}.  \label{DM}
\end{eqnarray}%
This expression illustrates that the condition $\partial ^{2}\ln B/\partial
z^{2}<0$, i.e. maximum of the magnetic field,     for $z=z_{m}$, is required for the existence of a smooth
(regular) solution.
Note that condition (\ref{DM}),   for $\partial^2 \ln B /\partial z^2 <0$,  also allows  the decelerating solution with $\partial M/\partial z <0 $. 

Equation (\ref{DM}) shows  that there are no regular solutions in case of the minimum of the magnetic field  suggesting the possibility of shock solutions. Various dissipative effects however may affect  the singularity and allow the flow with $\partial ^{2}\ln B /\partial z^{2} >0$ at the sonic point. For  example, ionization and charge-exchange modify the regularity condition as shown below in Section VI, Eq. (\ref{quadratic-eq-dM}).  More generally,  the  singularities may be resolved by collisional \cite{MirnovNF1972,SkovorodinPPR2012} and finite Debye length \cite{PekkerSovJPP1984} effects allowing the flows in complex non-monotonic magnetic field geometries. The finite Debye length effects may also result in formation of the double layers at the singular points.

\section{The accelerating potential for the case of  cold ions}

In this section we overview the case of cold ions providing  a simple illustration of the global properties of the accelerating potential formed by the converging-diverging magnetic field configuration.
For  $p_\perp=p_\parallel=0$, Eq.(\ref{gradient-of-M}) can be  integrated \cite{ManheimerIEEE2001,FruchtmanPoP2012} resulting
in the implicit equation for the ion velocity in the form%
\begin{equation}
\frac{M^{2}}{2}-\frac{1}{2}=-\ln \left( \frac{ B(z)}{M\left( z\right)B_{m}}%
\right).  \label{dl}
\end{equation}
Here $B_m=B(z)$ at  $z=z_m$ and the integration constant was chosen so the solution becomes regular at $z=z_m$ where $M=1$.

Generally,  acceleration of the magnetized plasma in the magnetic  nozzle is analogous to the gas flow in Laval nozzle. It is also similar to the problem  of the solar wind acceleration when the effective nozzle is created by the spherical expansion and the gravity force \cite{ParkerAJ1958}. 
The exact solution for equation (\ref{dl}) can be written \cite{SmolyakovPoP2021} in terms of the Lambert function which appears in  various fields including numerous plasma physics applications \cite{DubinovJPP2005,RaimbaultPoP2007,DubinovPoP2022}. The solution for spherically expanding solar wind can also be written with the Lambert function \cite{CranmerAJP2004}.

In the limit of $B(z)/B_0 \rightarrow 0$, from  Eq.(\ref{dl}) one can obtain two asymptotic solutions: $M\simeq B(z)/B_0$ and $M \simeq (-2\ln (B(z)/B_0))^{1/2}$. These asymptotics correspond to two branches of the regular solution, respectively,  in the converging and diverging parts of the nozzle. The singularity at $M=1$, where  the ion flow velocity is equal to $c_s$, can be removed if  $\partial \ln B/\partial z =0$. This condition corresponds to the smooth matching of two branches in Eq.(\ref{dl}) fixing the value of the velocity derivative at the sonic point.  Expanding  Eq. (13) near $M=1$  gives the expression for the velocity derivative $\partial M/\partial z$, cf. with Eq. (\ref{DM}) %
\begin{equation}
\left( \frac{\partial M}{\partial z}\right) ^{2}=-\left. \ \frac{1}{2}\frac{%
\partial ^{2}\ln B}{\partial z^{2}}\right\vert _{z=z_{m}}.  \label{dmdz}
\end{equation}

\begin{figure}[H]
    \centering
\subfloat[]{\includegraphics[width=0.45\linewidth]{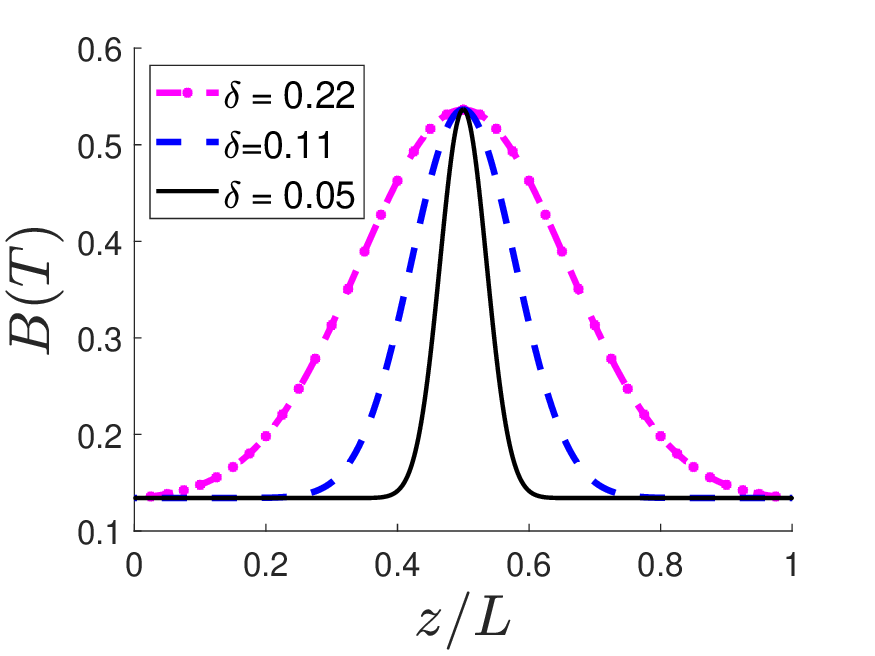}\label{cold-ions-different-B-fields-R-4}}
\subfloat[]{\includegraphics[width=0.45\linewidth]{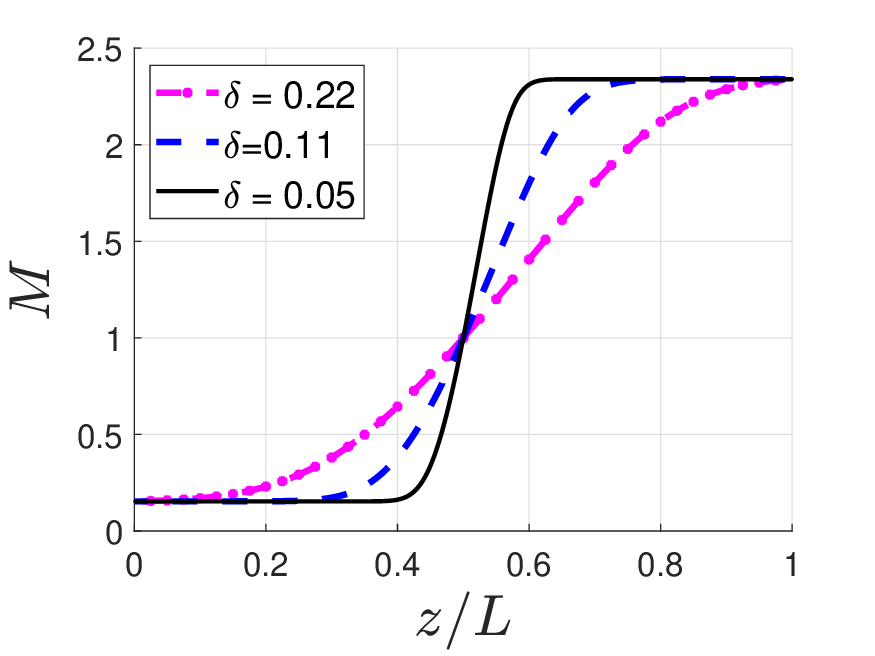}\label{cold-ions-M-R4}} \\
\subfloat[]{\includegraphics[width=0.45\linewidth]{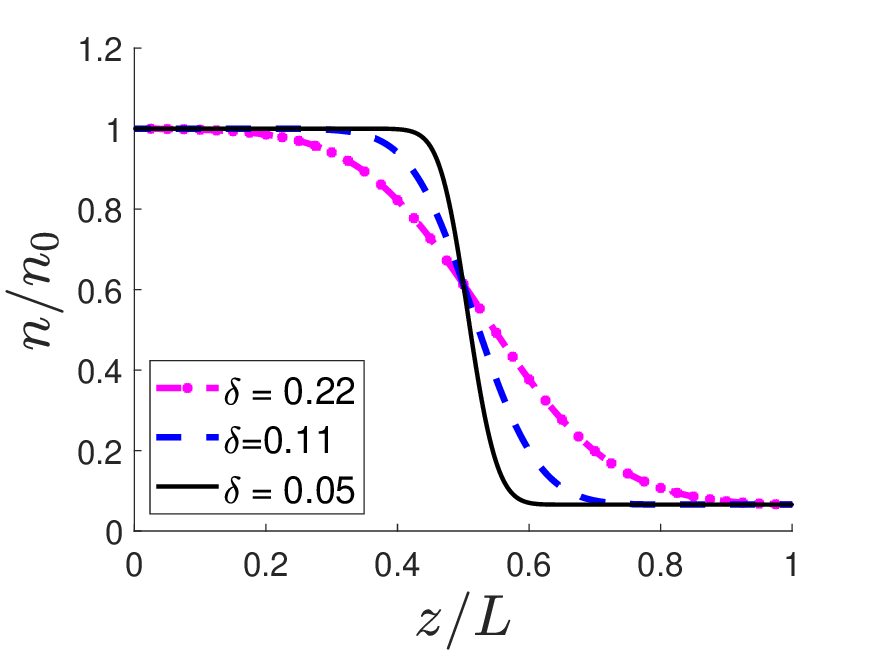}\label{cold-ions-plasma-density-R4}}
\subfloat[]{\includegraphics[width=0.45\linewidth]{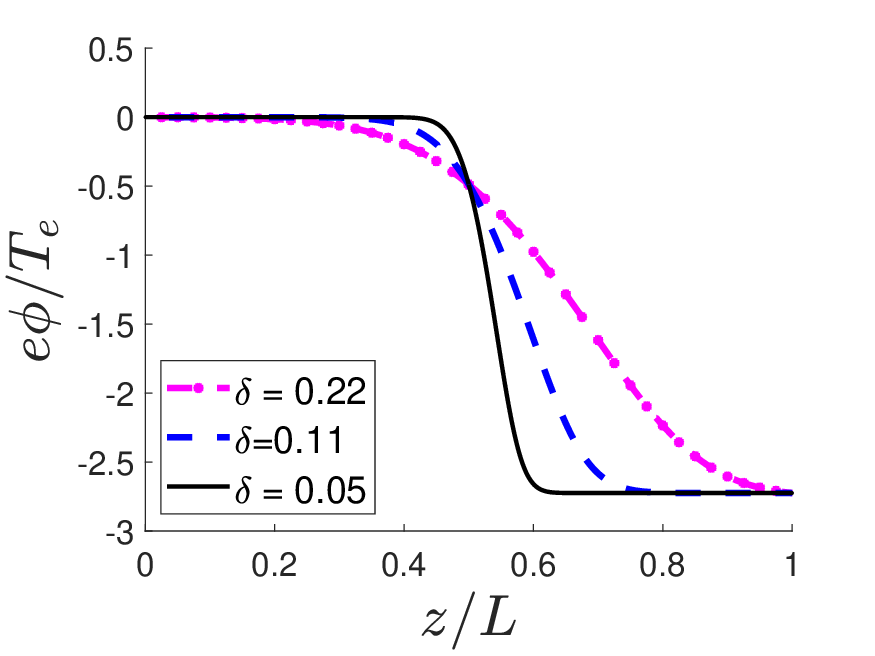}\label{cold-ions-potential-R4}}
\caption{Acceleration  of cold ions in the  magnetic field with mirror ratio $R=B_m/B(0) = 4$, $B(0)=B(L)$. The axial  profiles for a) magnetic field, b) Mach number, c) plasma density and d) electrostatic potential, for different values of the magnetic mirror width $\delta$.  }
\end{figure}

Here, for illustration, we consider the magnetic field mirror with  Gaussian profile, $B\left(z \right) = (B_{m}-B_{0}) \exp(-\left(z - z_0 \right)^2/(\delta^2 L^2)) + B_0$, giving a  mirror ratio $R =B_m/B(z)=4$ at both ends, $z=0$ and $z=L$. Several cases with different widths were considered by changing $\delta$, as shown in Fig. \ref{cold-ions-different-B-fields-R-4}. 

As it is shown in Figs.~1b-1c, changing the profile of the magnetic field under the constant mirror ratio modifies the velocity, density and potential profiles but the values at $z=0$ and $z=L$ remain the same. The plasma density is normalized to the value $n_0$ at $z=0$:   
Equation (\ref{dl}) shows that the local value of the plasma velocity is defined by the local value of the magnetic field. It is important to note that this solution has a global character defined by the regularization condition  at the sonic point, $V_\parallel =c_s$, which for cold plasma occurs at the point of the maximum magnetic field, where  ${\partial \ln B}/ {\partial z}=0 $.  The condition for the regular (smooth) solution  at the sonic point defines the velocity derivative at this point, therefore fixing the velocity profile globally. Changing the values of the mirror ratio $R$ modifies the velocities at $z=0$ and $z=L$ but the value at the maximum  $z=z_m$ remains equal to $c_s$, as shown in  Fig.~2. Note that  different values of the mirror ratio were considered by changing the value of both $\delta$ and $B_0$. The values of $\delta$ were adjusted as to ensure that the magnetic field with different R values would overlap at $z/L = 0.5$, i.e. to have the same derivative $\partial V_\Vert/\partial z$ at $z=z_m$. 
Increase of the mirror ratio leads to the decrease of the initial velocity $M_0\equiv M(0)<1$, and increase of the exit velocity at $z=L$, $M_L \equiv M(L)>1$. The plasma velocity at the nozzle exit only depends on the mirror ratio R (regardless of the details of the magnetic field profile), and has weak logarithmic divergence for $R>>1$.


\begin{figure}
\begin{center}
\includegraphics[width=80mm]{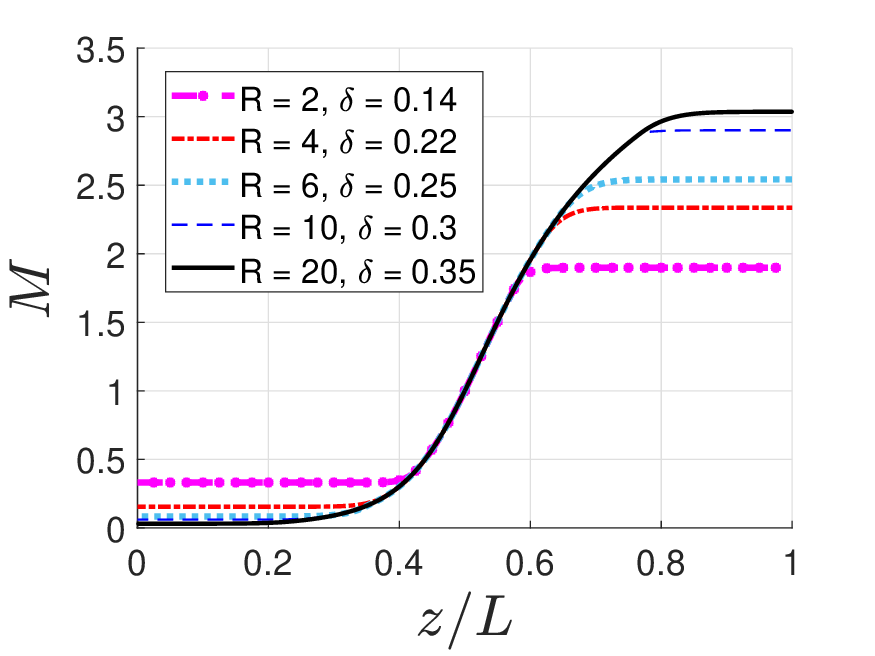}
\end{center}
\caption{Plasma velocity profiles for different values of the mirror ratio. }
\label{cold-ions-multiple-mirror-ratios}
\end{figure}




\section{Effect of the anisotropic ion temperature on plasma flow}

\begin{figure}
\begin{center}
\includegraphics[width=80mm]{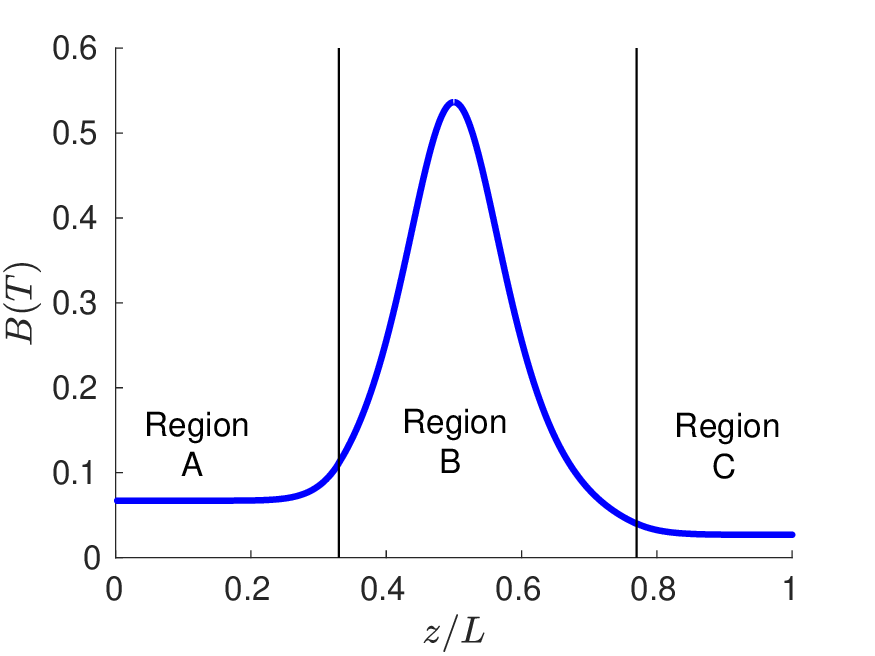}
\end{center}
\caption{Magnetic field of the mirror configuration as in Refs.  \onlinecite{onofri2017magnetohydrodynamic,BinderbauerAIP2016}}. 
\label{B-field}
\end{figure}

In this Section,  we present the global regular solutions for plasma flow  with finite (and anisotropic) ion pressure for the  magnetic field profile similar to the C-2U device \cite{onofri2017magnetohydrodynamic,BinderbauerAIP2016} shown in Fig. \ref{B-field}. The global solutions are considered in the region $0<z<L$.  The mirror ratios were $B_{m}/B_{0} = 8.0$ and $B_{m}/B_{r} = 20.0$ at the left, $z=0$, and the right, $z=L$,  ends of the nozzle, respectively.
The magnetic field setup is described in more detail in the Appendix. The left side of the simulation region represents the transition to the plasma source at $z<0$.

The magnetic geometry used in this paper considers only one side of  a typical magnetic mirror system like C-2U. This region has to be matched with the central source (plasma production and heating) region. Similar configurations are used in propulsion  applications with the mirror only on one side.  It is assumed in our study that the source region is at  $z < 0$. An important result of our study is the observation that plasma velocity is fixed globally by the condition at the singular point, so that the value of the velocity at $z=0$ cannot be arbitrary. The matching of the accelerating region (as studied here) with the source region is outside of the scope of this paper. However, some comments are provided in Section  VII. 

The numerical solution of stationary equations  
  (\ref{gradient-of-M}) and (\ref{potential-gradient} - \ref{p-perpendicular-with-charge-exchange}) are  obtained by the integration  from the the proximity of the sonic point, $z_{m}/L=0.5$, in both  $z<z_{m}$ and $z>z_{m}$ directions. The initial conditions for the parameters $M$, $n$, $p_{\parallel }$, and $p_{\perp }$ are obtained from the 
Taylor series expansion near $z=z_{m}$.  Across the entire nozzle electrons were assumed to be isothermal with $T_{e}=200\ eV$.  At the left end of the nozzle it was assumed that $T_{i\parallel
_{0}}=T_{i\perp _{0}}=200\ eV$ with the 
density  $n_{0}=1.0\times 10^{19}\ m^{-3}$,  however all results can be represented
in dimensionless units by being normalized to their respective values at $z = 0$. 
The shooting method
was used to achieve the final solution with isotropic ion pressure, $T_{i}{}_{\parallel }{}_{0}=T_{i}{}_{\perp }{}_{0}=200\ eV$ at the
left end of the nozzle, which corresponds to the beginning of Region A in Fig. \ref{B-field}.

Similarly to the case of cold ions, plasma acceleration occurs in the regions with the finite gradient of the magnetic field.  The perpendicular ion pressure enhances the ion acceleration according to equation (\ref{DM}). The parallel ion pressure shifts the sonic point which is now defined by Eq. (10). However, the parallel pressure decreases fast with distance, so the effects of the finite parallel pressure on the location of the singular point is weak for our parameters.

\begin{figure}[H]
    \centering
\subfloat[]{\includegraphics[width=0.45\linewidth]{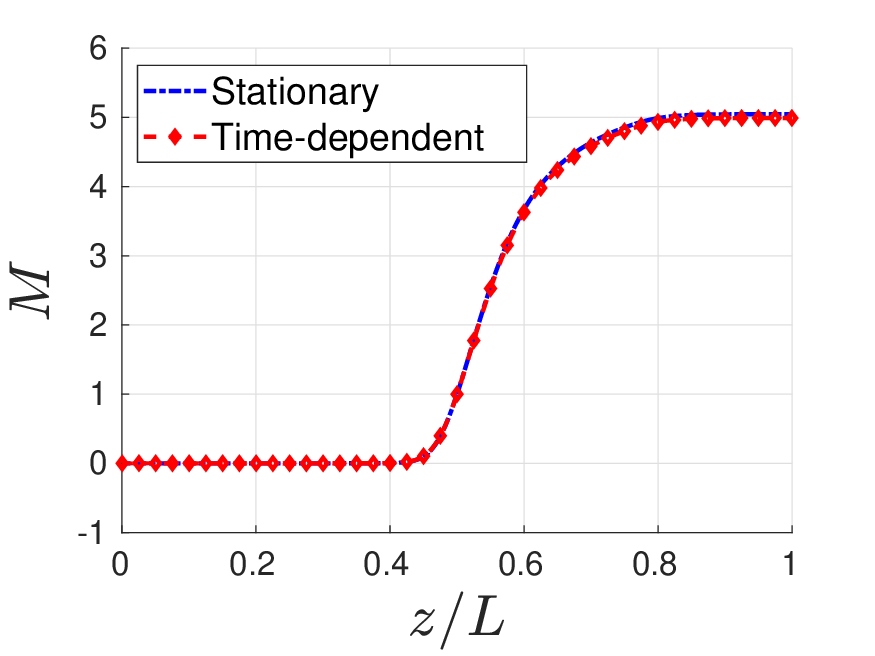}\label{M-stationary}}
\subfloat[]{\includegraphics[width=0.45\linewidth]{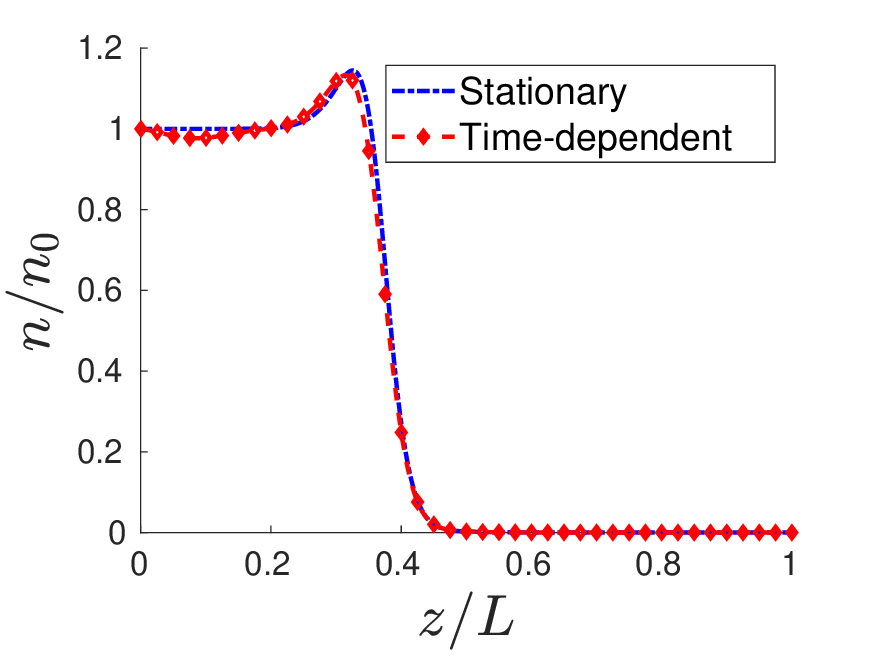}\label{n-stationary}} \\
\subfloat[]{\includegraphics[width=0.45\linewidth]{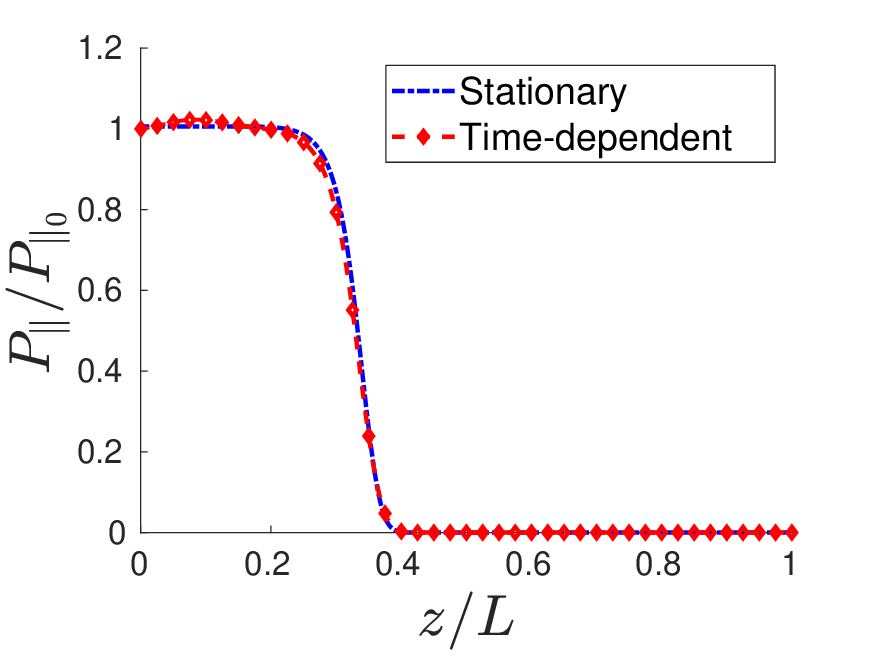}\label{p-parallel-stationary}}
\subfloat[]{\includegraphics[width=0.45\linewidth]{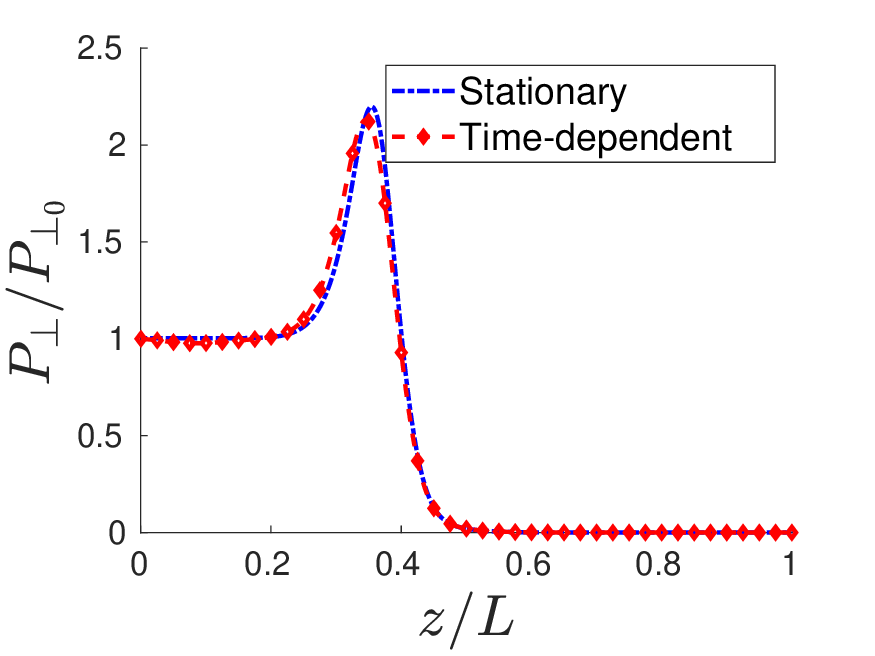}\label{p-perpendicular-stationary}} \\
\subfloat[]{\includegraphics[width=0.45\linewidth]{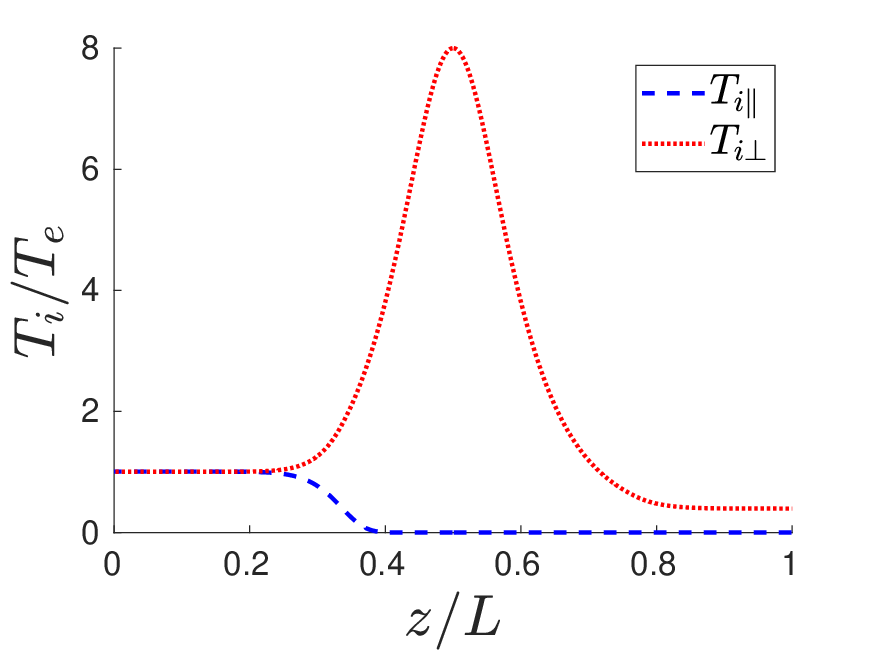}\label{T_i}} 
\subfloat[]{\includegraphics[width=0.45\linewidth]{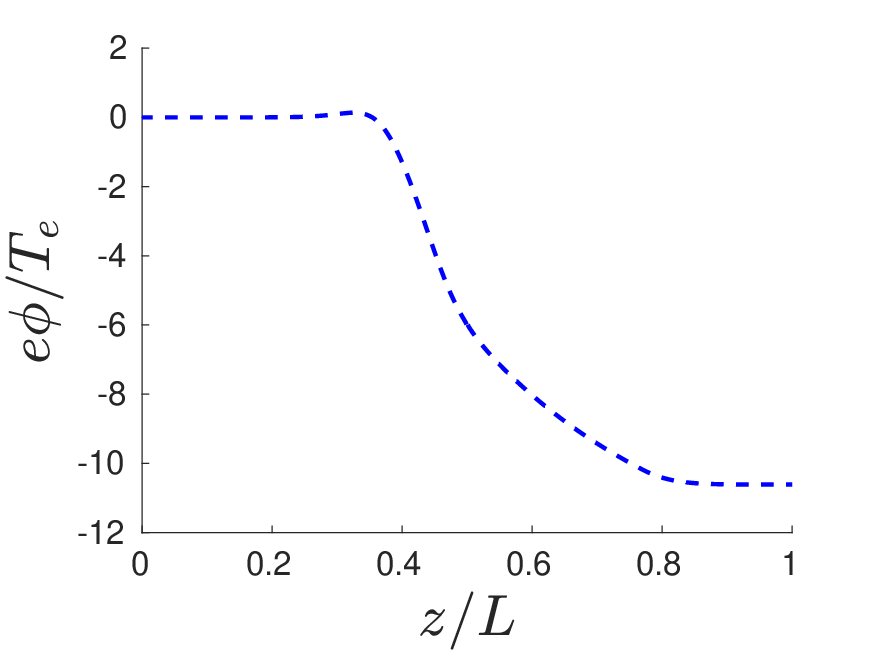}\label{potential}}
    \caption{Stationary solutions profiles for the case of $T_{i\parallel0}=T_{i\perp0}= 200\ eV$: a) Mach number, b) plasma density, c) parallel pressure, d) perpendicular pressure, e) ion temperatures and f) electrostatic potential. Both solutions of stationary equations (obtained by shooting methods) and time-dependent initial value problem are shown in (a), (b),(c) and (d).}
\end{figure}

An interesting observation that can be seen from Fig. \ref{n-stationary} is that for a finite ion temperature  there is a region where the density is increasing with distance, contrary to 
the case of cold ions, where the density is always monotonically decreasing in regions of non-constant magnetic field. A similar behavior is observed for the potential, with the small increase with distance being displayed in Fig. \ref{potential}. \\ 
\ 
The perpendicular plasma pressure follows the increase of the magnetic field $B$ related to the conservation of the adiabatic invariant as $
T_{i\perp}(z)/B(z) = const$. The  dependence of $T_{i\perp}, T_{i\Vert}$ on $B(z)$ are shown in Fig. \ref{T_i}. The  decrease of $T_{i\Vert}$  in the outer region is related to strong plasma density dependence $ ~n^3$ in the CGL equation of state for $p_\Vert$ and density decrease due to plasma acceleration.

\begin{figure}
\begin{center}
\includegraphics[width=80mm]{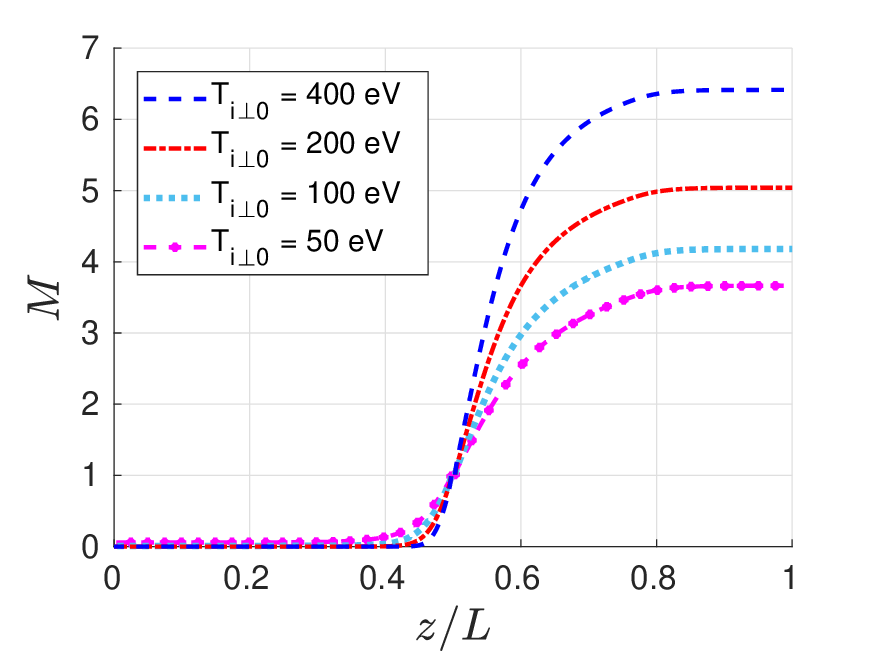}
\end{center}
\caption{Mach number M profiles for $T_{i\parallel0} = 200\ eV$ and different $T_{i\perp0}$ values.}
\label{M-different-ion-temperatures}
\end{figure}

The perpendicular ion temperatures $T_{i\perp0} = 200\ eV$ were changed
to study the effect of perpendicular temperature at the nozzle entrance $z=0$. Three
additional cases were studies with $T_{i\perp} = 400\ eV$, $T_{i\perp} = 100\ eV$, and $T_{i\perp} = 50\ eV$, while $T_{i\parallel0} = 200\ eV$ and $T_e = 200\ eV$ remained the same. For the initial anisotropic ion pressure state at $z=0$,  the finite perpendicular ion temperature increases the velocity of the accelerated ions as shown in Fig. \ref{M-different-ion-temperatures} and displayed in Table I.

Plasma temperatures can be lower in some laboratory devices \cite{MakrinichPoP2009}, e.g. $T_e = 10 \ to \ 30 \ eV$ and $T_i = 0.5 \ to \  3 \ eV$. The ion temperatures with $T_i \ll T_e$ will  resemble the profiles of the cold ion case.
 Since, the speed of sound $c_s$ has been normalized as $c_s = \sqrt{T_e/m_i}$, the lower values of $T_e$ will result in a lower absolute values of the velocities. The ion temperatures profiles will qualitatively remain similar, e.g. as in Fig. 4.

\begin{table}
\begin{center}
\begin{tabular}{ |c|c|c|c|} 
\hline
$T_{i\parallel0} (eV)$ & $T_{i\perp0} (eV)$ & $M_0$ & $M_L$  \\ 
\hline
 200 & 400 & $2.85 \times 10^{-7}$ & 6.41 \\ 
 200 & 200 & $3.16 \times 10^{-4}$ & 5.04 \\ 
 200 &  100 & 0.01               & 4.18 \\ 
 200 &   50 & 0.06               & 3.66 \\ 
\hline
\end{tabular}
\caption{Plasma flow Mach number for different ion temperatures at the nozzle entrance, $z=0$, and exit, $z=L$. }
\end{center}
\end{table}

\begin{figure}
\begin{center}
\includegraphics[width=80mm]{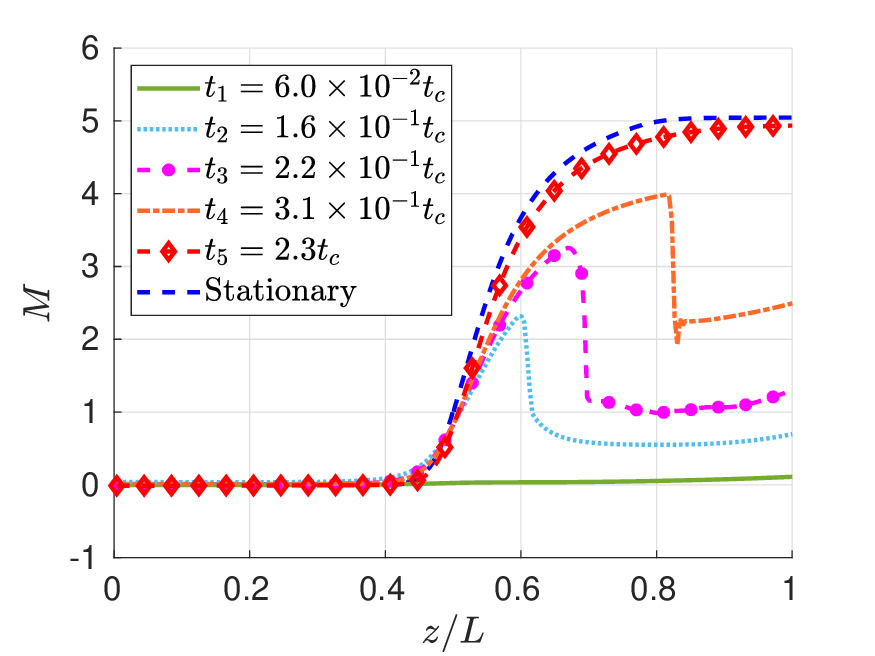}
\end{center}
\caption{Time evolution of  the velocity profile obtained in the time dependent initial value problem. }
\label{M-time-evolution}
\end{figure}

 Steady-state solutions of stationary equations  obtained by the shooting method were compared
 with the time-dependent solutions obtained as a solution of the initial value problem for equations (\ref{n-time-dependent-with-charge-exchange} - \ref{p-perpendicular-with-charge-exchange}), presented in the Appendix with additional details. We have verified that the
solution of the time-dependent equations converge well to the stationary solution obtained by the shooting  method, as shown in figures \ref{M-stationary}, \ref{n-stationary}, \ref{p-parallel-stationary} and \ref{p-perpendicular-stationary}. The characteristic time was $t_c = L/c_s = 4.0 \times 10^{-5} s$ and each time interval was given as a multiple of $t_c$. Each time step in the simulation was equivalent to $1.0 \times 10^{-5} t_c$. The evolution of the time-dependent problem towards the stationary solution is shown in Fig. \ref{M-time-evolution}  for the successive moments in time  $t_1 < t_2 ... < t_5$ with the lowest $t_1 = 6.0 \times 10^{-2}t_c$ and the stationary solution value being reached at $t=2.4t_c$.
This exercise demonstrates that the stationary accelerating solutions are stable and therefore dynamically accessible in the initial value problem.

\section{Ionization and Charge-Exchange Effects}

In this section we consider the ion flow taking into account ionization and charge-exchange effects. These effects are important in fusion applications due to presence of neutrals in the mirror region. Effects of neutrals were also considered in  propulsion applications \cite{MakrinichPoP2009,FruchtmanJPhysD2017,FruchtmanIEEE2011}. 
Depending on the application, neutral density and therefore ionization   and charge-exchange coefficients may vary in a rather wide range    \cite{lieberman2005principles,onofri2017magnetohydrodynamic,BinderbauerAIP2016,MakrinichPoP2009}.
 To estimate possible effects at various conditions,  here we use a generic form of the sink terms $S_\parallel$ and $S_\perp$ with various values of $\nu _{1}$ and $\nu_{2}$ in normalized form 
$\nu_2^{\prime }=\nu _{2}L/c_{s}$,  $\nu _{1}^{\prime }=\nu _{1}L/c_{s}$ from low to large values. We have used  the values of  $L = 4 \ m$ as the length of the nozzle and $c_s = 9.78 \times 10^4 \  m/s$ as the speed of sound \cite{BinderbauerAIP2016} giving for  $\nu'_1 \in [10^{-2}.. 0.5]$ and  $\nu'_2 \in [0.1..2]$  with the dimensional values of $\nu_1 \in [245 \ s^{-1} .. 1.22 \times 10^4 \ s^{-1}]$ and  $\nu_2 \in [2.45 \times 10^3 \ s^{-1}..4.9 \times 10^4 \ s^{-1}]$, respectively. The ionization and charge-exchange coefficients are   $\nu_1 = \beta_{iz}N_g$ and $\nu_2 = \beta_{cx} N_g$,  where $N_g$ is the density of neutral atoms, and typical rates \cite{janev1987collision} $\beta_{iz} = 3.1 \  \times 10^{-14} \ m^3 \ s^{-1}$ and $\beta_{cx} =  6.18 \ \times 10^-{14} \ m^3 \ s^{-1}$.   For our parameters above, it gives for the values in the range   $\nu_1 \in [245 \ s^{-1}.. 1.22 \times 10^4 \ s^{-1}]$ the neutral density  $N_g \in [7.8 \times 10^{15} \  m^{-3}.. 3.8 \times 10^{17} \ m^{-3}]$, and for $\nu_2 \in [2.45 \times 10^3 \ s^{-1}... 4.9 \times 10^4 \ s^{-1}]$ -- the values $N_g \in [4.0 \times 10^{16} \  m^{-3}.. 7.9 \times 10^{17} \ m^{-3}]$. Note that at lower boundaries of these values the collisional effects are not significant. Therefore, our parameters cover a wide range of typical densities observed in experiments \cite{BinderbauerAIP2016} and also extend to the low collisionalities typical for space propulsion applications. It should be noted however that  generally in applications the density of neutral atoms is not constant while we assumed the constant and uniform profile. Therefore, our results should be understood as a parametric study towards revealing the main trends.

For the applications in Ref.~\onlinecite{BinderbauerAIP2016} the lowest value of  the ion cyclotron frequency is $\omega_{ci} = 7.3 \times 10^6 \ rad \ s^{-1}$ remaining larger than the collisional frequencies, $\omega_{ci} >> (\nu_1,\nu_{2}),$ therefore justifying the use of the MHD approximation and assumption of magnetized ions. The problem of detachment \cite{HooperAIAA1993,BreizmanPoP2008} that occurs for space propulsion applications with lower values of the magnetic field  is not considered in our paper.

In dimensionless form the stationary equations describing ion dynamics have the form
\begin{equation}
\frac{\partial n}{\partial z}=n\left(\frac{\partial\ln B}{\partial z}-\frac{%
1}{M}\frac{\partial M}{\partial z}+\frac{\nu_1}{M}\right),
\label{gradient-of-n-vion}
\end{equation}
\begin{equation}
\frac{\partial p_\parallel}{\partial z}=p_\parallel\left(\frac{\partial\ln B%
}{\partial z}-\frac{3}{M}\frac{\partial M}{\partial z}-\frac{\nu_2}{M}%
\right),  \label{gradient-of-p-para-vion}
\end{equation}
\begin{equation}
\frac{\partial p_\perp}{\partial z}=p_\perp\left(2\frac{\partial\ln B}{%
\partial z} -\frac{3}{M}\frac{\partial M}{\partial z}-\frac{\nu_2}{M}%
\right),  \label{gradient-of-p-per-vion}
\end{equation}
\begin{equation}
\begin{split}
\left( M^{2}-1-\frac{3p_{\parallel}}{n}\frac{T_{i\parallel0}}{T_{e}}\right) \frac{\partial M}{%
\partial z}=& -\left( 1+\frac{p_{\perp }}{n}\frac{T_{i\perp0}}{T_{e}}\right) M\frac{\partial
\ln B}{\partial z} \\
& +\left(\frac{p_{\parallel}}{n}\frac{T_{i\parallel0}}{T_{e}}-M^{2}\right) \nu_{2}-\nu_{1},
\end{split}
\label{gradient-of-M-vion}
\end{equation}
\begin{equation}
\phi= T_e \ln\left(\frac{n}{n_{0}}\right).
\label{gradient-of-phi}
\end{equation}
 The plasma parameters here are normalized to their respective values
at the left end of the nozzle such that 
$n^{\prime }=n/n_{0}$, $%
p_{\parallel }^{\prime }=p_{\parallel }/p_{\parallel_0}$, $%
p_{\perp }^{\prime }=p_{\perp }/p_{\perp_0}$, $T^{\prime
}_{i_\parallel }=T_{i_\parallel }/T_{e}$, $T^{\prime
}_{i_\perp }=T_{i_\perp }/T_{e}$, $z^{\prime }=z/L$ and $%
t^{\prime }=c_{s}t/L$. For the
sake of convenience all the primes on parameters will be dropped and it will be assumed that n, $p_{\parallel }$, $p_{\perp }$, $%
T_{i}{}_{\parallel }$, $T_{i}{}_{\perp }$, $\nu_{1}$, $\nu_{2}$, $z$ and $t$ represent normalized quantities.

Additional terms in Eq.~(\ref{gradient-of-M-vion}) due to the ionization and charge-exchange effects modify the regularization condition
at the sonic point. The sonic point is still defined by condition \ref{M-sonic-point}. However the location of the sonic point, where the right hand side of equation (\ref{gradient-of-M-vion}) is zero, is shifted from the position of the maximum magnetic field at $z=z_m.$   
Expanding near the point, where both sides of Eq. (\ref{gradient-of-M-vion}) equal zero, we obtain for $\partial M/\partial z$ the following equation
\begin{equation}
a\left(\frac{\partial M}{\partial z}\right)^2 + b\left(\frac{\partial M}{%
\partial z}\right) + c = 0,  \label{quadratic-eq-dM}
\end{equation}
where 
\begin{equation}
a=2\left(1+6 T_{i\parallel S}\frac{T_{i\parallel 0}}{T_e}\right),  
\label{a}
\end{equation}

\begin{equation}
\begin{split}
b = \ & 5T_{i\parallel S}\frac{T_{i\parallel 0}}{T_e}\nu_2 + 3T_{i\parallel S}\frac{T_{i\parallel 0}}{T_e}%
\nu_1 +\left(1+T_{i\perp S}\frac{T_{i\perp 0}}{T_e}\right)\left(1+3T_{i\parallel S}\frac{T_{i\parallel 0}%
}{T_e}\right)^\frac{1}{2} \\
& \times \frac{\partial\ln B}{\partial z}+2\left(1+3T_{i\parallel S}\frac{T_{i\parallel 0}}{%
T_e}\right)\nu_2,
\end{split}
\label{b}
\end{equation}
\begin{equation}
\begin{split}
c = & \left(1+T_{i\perp S}\frac{T_{i\perp 0}}{T_e}\right)\left(1+3T_{i\parallel S}\frac{T_{i\parallel 0}%
}{T_e}\right)\frac{\partial^2 \ln B}{\partial z^2}+T_{i\perp S}\frac{T_{i\perp 0}}{T_e}
\\
& \times \left(1+3T_{i\parallel S}\frac{T_{i\parallel 0}}{T_e}\right) \left(\frac{%
\partial\ln B}{\partial z}\right)^2 -T_{i\perp S}\frac{T_{i\perp 0}}{T_e}\left(1+3T_{i\parallel S}\frac{%
T_{i\parallel 0}}{T_e}\right)^\frac{1}{2} \\
& \times \frac{\partial\ln B}{\partial z}\left(\nu_2+\nu_1\right) +T_{i\parallel S}\frac{%
T_{i\parallel 0}}{T_e}\left(\nu_2+\nu_1\right)\nu_2.
\end{split}
\label{c}
\end{equation}
Here  $T_{i\parallel S}$ and $T_{i\perp S}$ are the normalized values of the parallel
and perpendicular temperatures at the sonic point, which no longer occurs at $%
z=z_m$ but is shifted to the right. The shift in the location of the sonic is directly affected by the value of $p_\parallel$ that is in turn affected by the values of $\nu_1$ and $\nu_2$. Thus, higher values of $\nu_1$ and $\nu_2$ will have a more profound effect on $p_\parallel$ and will indirectly shift the sonic point further away from its initial position at $z=z_m$.
Similar to Section V,  we obtain 
stationary solutions for equations (\ref{gradient-of-n-vion}) - (\ref{gradient-of-phi}) using the shooting
method as described in Section III and compare  these solutions with the solution of the initial value problem given by equations (\ref{potential-gradient}) - (\ref{p-perpendicular-with-charge-exchange}).
Again, it has been confirmed that time-dependent solutions converge well to the
stationary solutions after some relaxation.   

\begin{figure}[H]
    \centering
\subfloat[]{\includegraphics[width=0.45\linewidth]{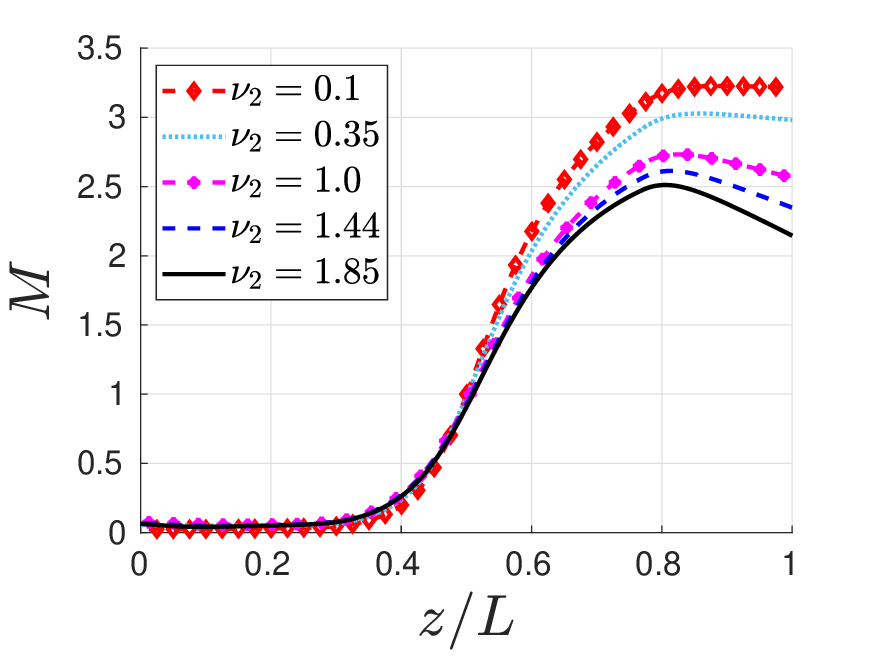}\label{M-multiple-charge-exchange}}
\subfloat[]{\includegraphics[width=0.45\linewidth]{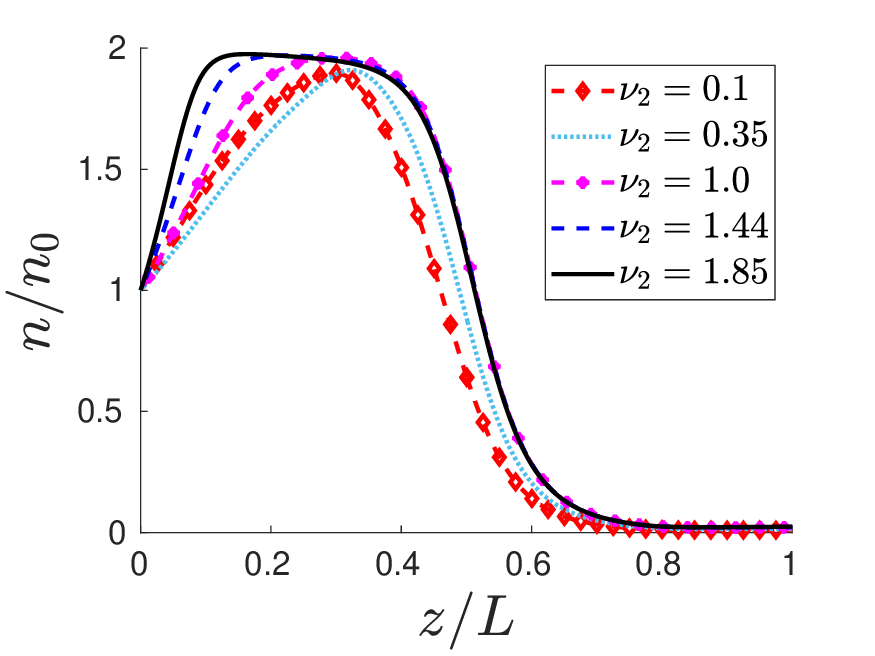}\label{plasma-density-multiple-charge-exchange}} \\ 
\subfloat[]{\includegraphics[width=0.45\linewidth]{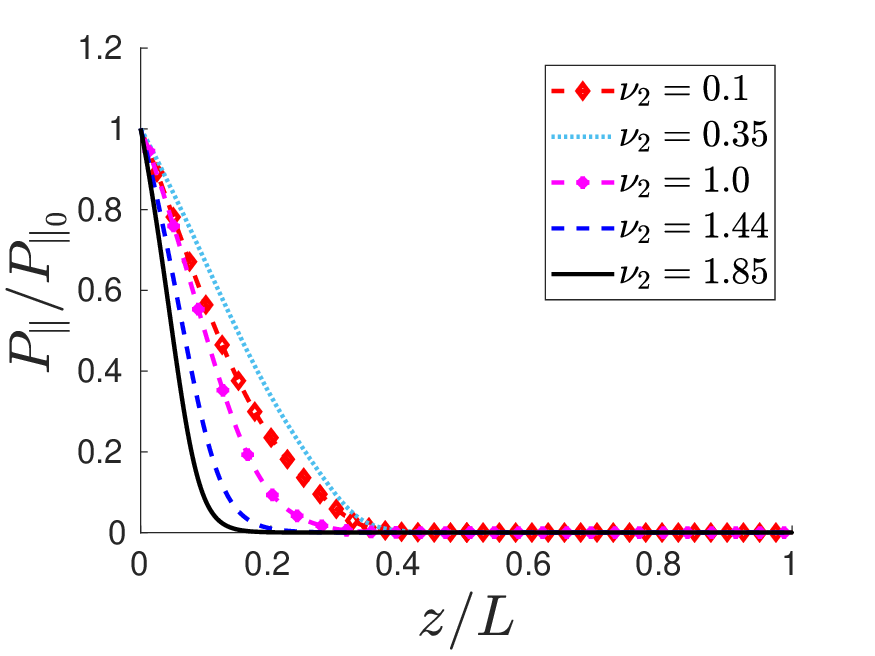}\label{p-parallel-multiple-charge-exchange}}
\subfloat[]{\includegraphics[width=0.45\linewidth]{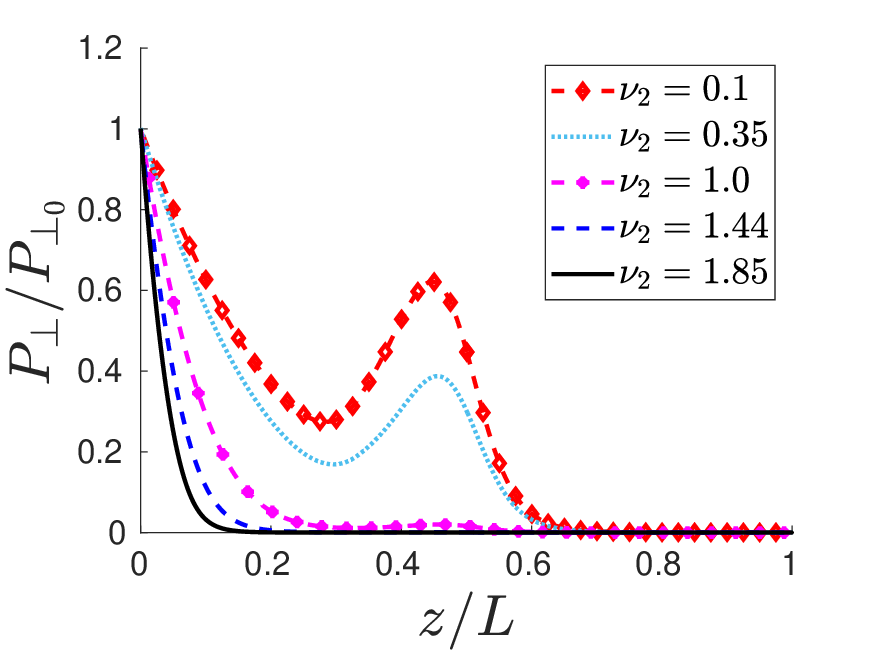}\label{p-perpendicular-multiple-charge-exchange}} \\ 
\subfloat[]{\includegraphics[width=0.45\linewidth]{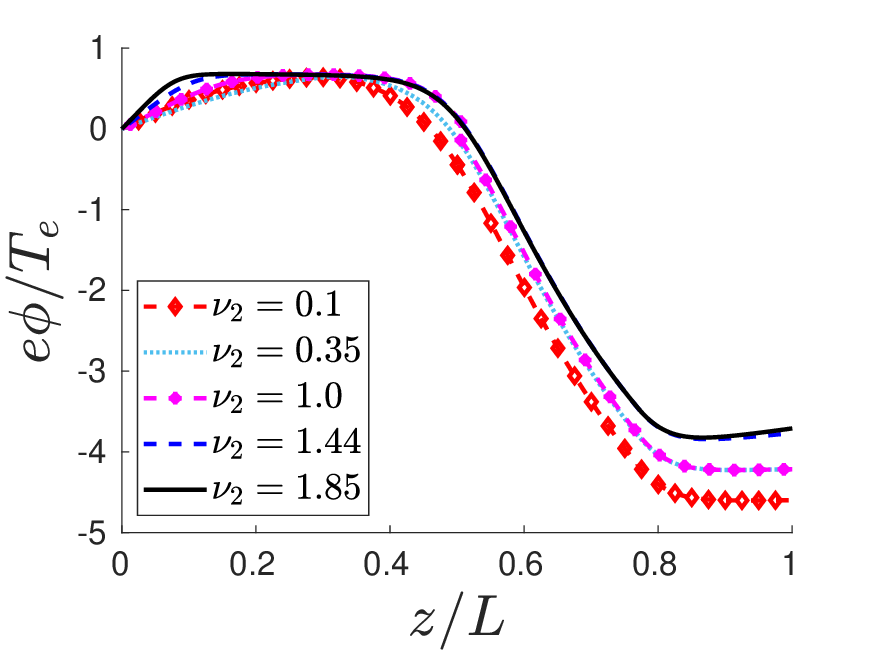}\label{potential-multiple-charge-exchange}}
    \caption{Stationary solutions for the cases with constant ionization $\nu_{1} = 0.1$ and different values of charge-exchange $\nu_{2}$ for a) Mach number, b) plasma density, c) parallel pressure, d) perpendicular pressure and e) electrostatic potential.}
\end{figure}

\begin{figure}[H]
    \centering
\subfloat[]{\includegraphics[width=0.45\linewidth]{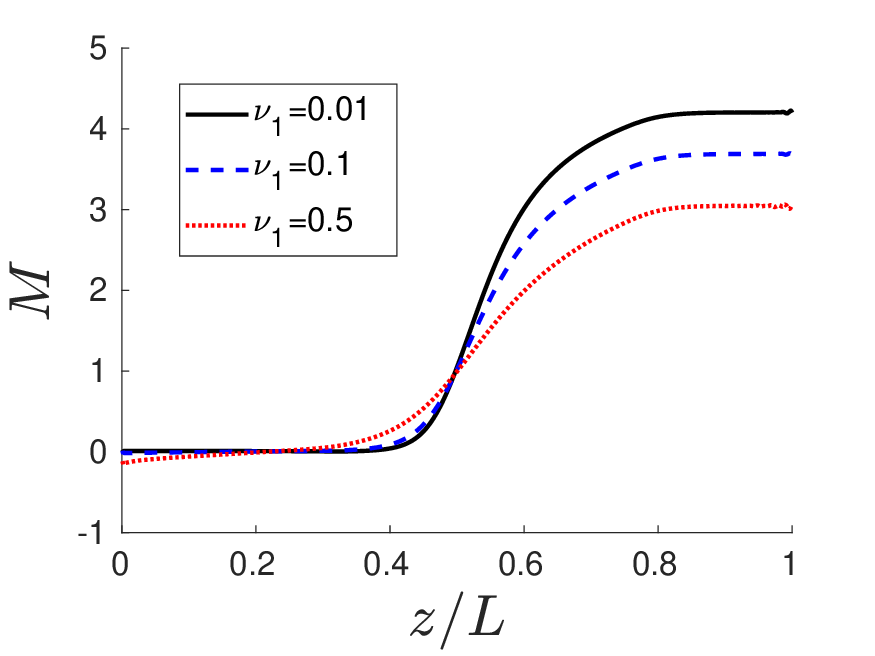}
\label{mach_number_multiple_ionization}}
\subfloat[]{\includegraphics[width=0.45\linewidth]{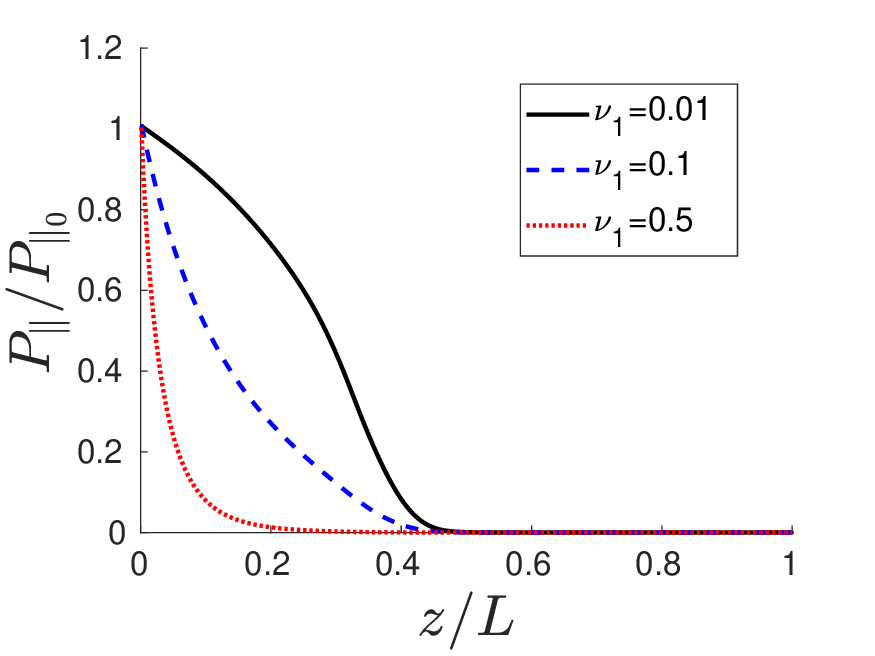}
\label{parallel_pressure_multiple_ionization}} \\
\subfloat[]{\includegraphics[width=0.45\linewidth]{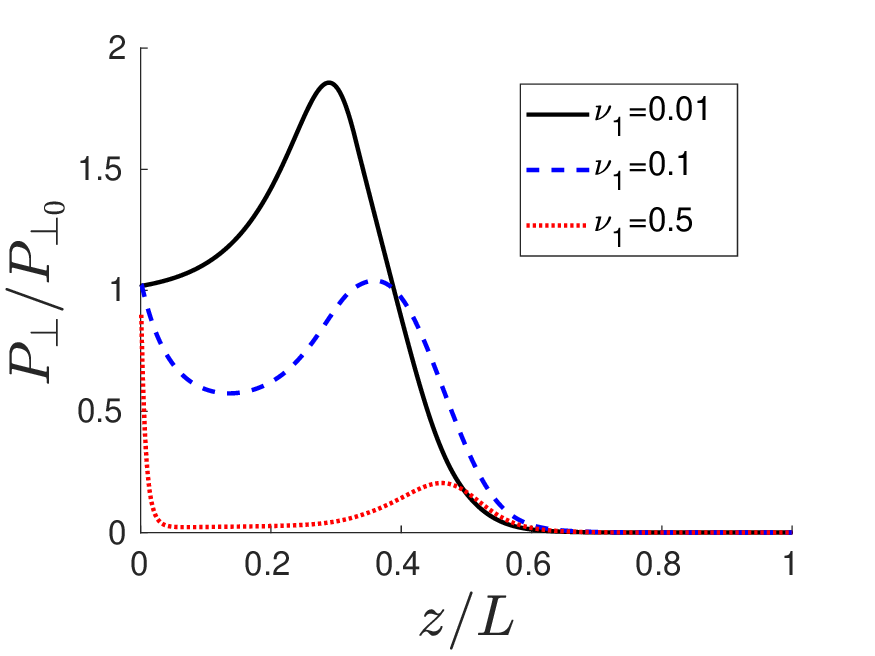}
\label{perpendicular_pressure_multiple_ionization.eps}}
\subfloat[]{\includegraphics[width=0.45\linewidth]{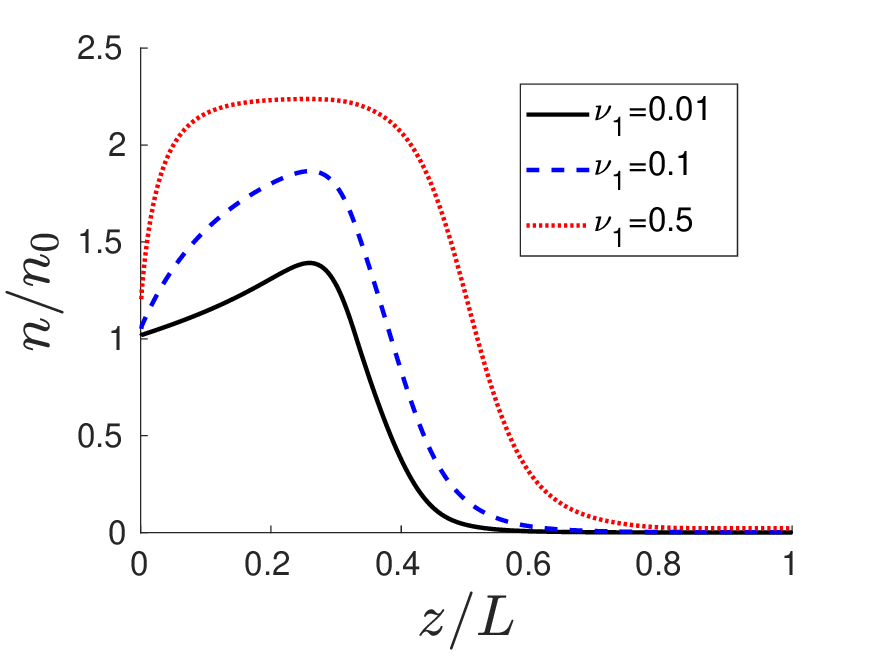}
\label{plasma_density_multiple_ionization.eps}}
    \caption{Stationary solutions for the cases with constant charge-exchange $\nu_{2} = 0$ and different values of ionization $\nu_{1}$: a) Mach number, b) parallel pressure, c) perpendicular pressure and d) plasma density.}
\end{figure}

There are several features introduced by ionization and charge-exchange effects. 
One is a non-monotonous behavior of plasma density with the increase 
due to the ionization, which is especially noticeable  in  the region of flat magnetic field before any substantial acceleration occurs. This behavior is more pronounced for higher ionization values, as indicated in Fig. \ref{plasma_density_multiple_ionization.eps}. Related to the density behavior, the potential also shows a non-monotonous  increase in the region to the left of the maximum of the magnetic field as shown in Fig. \ref{potential-multiple-charge-exchange}. 

Another effect introduced by the dissipative terms, is the modification of the perpendicular pressure profile so that it does not follow the increase in the magnetic field. The latter effect is much reduced by dissipation so the perpendicular pressure profile may become similar to the parallel pressure with almost monotonous decrease throughout the whole region, and  both the
parallel and perpendicular pressure having very similar profiles with  $
T_{i_\parallel}$ and $T_{i_\perp}$ as displayed in Figures \ref{p-parallel-multiple-charge-exchange} and \ref{p-perpendicular-multiple-charge-exchange}.  
With the above noted modifications,  the resulting velocity has a significantly larger value at the left boundary,  $M = 0.127$, compared  with the case in the absence of charge-exchange and ionization, and a lower final acceleration value on  the exit side of the nozzle. The reduction in the final value of acceleration is greater for higher charge-exchange values as shown in Fig. \ref{M-multiple-charge-exchange} but also in the case of ionization alone as shown in Fig. \ref{mach_number_multiple_ionization}.  One has to note also that the presence of the dissipative terms due to the ionization and charge-exchange results in the shift of the position of the sonic point so it is no longer at the magnetic field maximum, as described by equation (\ref{quadratic-eq-dM}).

The effect  of charge-exchange has been further studied for various values of $\nu_2$ as shown in Table II and Figures \ref{M-multiple-charge-exchange}, \ref{plasma-density-multiple-charge-exchange}, \ref{p-parallel-multiple-charge-exchange}, \ref{p-perpendicular-multiple-charge-exchange} and \ref{potential-multiple-charge-exchange}. In Table II, the values of $M_{0}$ and $M_{L}$ represent the value of the Mach number at the $z/L = 0$ and $z/L = 1$ ends of the nozzle respectively. Further increase in $\nu_2$ results in the non-monotonous behavior of the plasma velocity in the outer region, namely Region C, where the magnetic field is almost flat, between $z/L=0.8$ and $z/L=1$, see Fig. \ref{M-multiple-charge-exchange}. The width of the region with the density increase on the left side of the mirror is also increasing with $\nu_2$ as shown in Fig. \ref{plasma-density-multiple-charge-exchange}. The characteristic increase in the perpendicular pressure so well pronounced for $\nu_2=\nu_1=0$, is smoothed out by finite values of  $\nu_2$ and eventually disappears for large values of $\nu_2$ as shown in Fig. \ref{p-perpendicular-multiple-charge-exchange}. A similar behavior is observed for $p_{\parallel}$ and $p_{\perp}$ when only ionization is present, with the decrease being more pronounced for higher values of $\nu_1$ as shown in Fig. \ref{parallel_pressure_multiple_ionization} and Fig. \ref{perpendicular_pressure_multiple_ionization.eps} respectively. An interesting observation is that the density curve corresponding to $\nu_2 = 0.35$, is below the curve corresponding to $\nu_2 = 0.1$ in the Region A where the magnetic field is constant. A larger charge-exchange value would result in a greater drag force on the ions, reducing the Mach number and resulting in a higher plasma density. It could be possible that when the rate of ionization is equal to the rate of charge-exchange, as is the case for $\nu_1 = \nu_2 = 0.1$, the neutrals that formed in the region of constant magnetic field as a result of charge-exchange are themselves ionized, thus enhancing ionization and resulting in a greater increase in plasma density. The same effect is seen for the parallel pressure in Fig. \ref{p-parallel-multiple-charge-exchange} and this effect is likely a result of the behavior of the plasma density.
 The reduction of the velocity at the exit side, corresponding to the end of Region C, is consistent  with the lower overall drop in the electrostatic potential as shown in Fig. \ref{potential-multiple-charge-exchange} and displayed in Table II.   
 
\begin{table}[H]
\begin{center}
\begin{tabular}{ |c|c|c|c|c|} 
\hline
 $\nu_2$ & $M_0$                 & $M_L$   & $\left({e\phi/T_e}\right)_0$ & $\left({e\phi/T_e}\right)_L$  \\ 
 \hline
 0          & $7.5 \times 10^{-3}$  & 3.32  & $0$             & -4.77 \\ 
 0.1        & 0.024                & 3.217   & $0$             & -4.6 \\ 
 0.35       & 0.054                & 2.981   & $0$             & -4.21 \\ 
 1.0        & 0.071               & 2.564   & $0$             & -3.83 \\ 
 1.44       & 0.077               & 2.348   & $0$             & -3.77 \\ 
 1.85       & 0.127               & 2.145   & $0$             & -3.71 \\ 
\hline
\end{tabular}
\caption{ Plasma Mach number and electrostatic potential  for constant ionization $\nu_{1} = 0.1$ and different values of charge-exchange $\nu_2$.}
\end{center}
\end{table}
 
\section{Discussion and conclusions}

The flow of plasma in a magnetic nozzle was studied in the paraxial approximation using  fluid MHD model with fully magnetized ions and anisotropic pressure.  Stiffness  of the global acceleration profile,  which is fully defined by the  regularization condition at the sonic point, is one of the most important results of this manuscript.  The unique accelerating solution  starts with  a finite (and fixed) value $V_\Vert<c_s$  at the entrance of the magnetic mirror. This solution continues  through the sonic point  $V_\Vert=c_s$ to the diverging region  where the ion velocity becomes supersonic. Plasma acceleration occurs as a result of an ambipolar electrostatic potential drop formed in the magnetic mirror, converting finite plasma pressure into the ion kinetic energy. We have shown that the finite perpendicular ion temperature increases the finite ion velocity in the exit region due to the effect of the mirror force. The finite (accelerated) value of the plasma velocity is defined by the electron and ion temperatures and the magnetic mirror  aspect ratio. 
It is important to note that CGL theory is collisionless and the additional ion acceleration due to the perpendicular pressure is a result of the mirror force.

In many applications plasmas are collisionless but coupling of plasma flow with neutrals was noted as an important factor for some fusion \cite{onofri2017magnetohydrodynamic,NgPoP2007} and plasma thruster applications \cite{FruchtmanJPhysD2017,WachsPSST2020,TakahashiAPL2016}.
We have studied how the ionization and charge-exchange processes reform the accelerating potential impeding the plasma flow. We have shown that the ionization and charge-exchange  shift the position of the sonic point from the maximum magnetic field, reduce  plasma acceleration,  and may result in non-monotonous behavior of plasma density and electrostatic potential. 

We have obtained solutions of stationary equations with the shooting method. We have also performed simulations of time-dependent equations as an  initial value  problem therefore proving the stability of the obtained stationary solutions. 
 
 Similarly to Refs. \onlinecite{MerinoPSST2018,AhedoPSST2020}, our model is a paraxial approximation which considers the radial variations only in the main order. Though such one-dimensional models are often a good approximation  for a full two-dimensional dynamics \cite{FruchtmanPoP2012},  two-dimensional effects are expected to be important \cite{AhedoPoP2010} for large diameter systems. 

In general ion kinetic effects are important for low density collisionless plasma as  in Refs. \onlinecite{BinderbauerAIP2016,onofri2017magnetohydrodynamic}.  One of the related limitations of the current study is the use of  the standard collisionless  two-pressure CGL model which neglects the ion  heat fluxes. Further work is required to include heat fluxes either with extended two-pressure models and/or with full kinetic theory \cite{AhedoPSST2020,MartinezPoP2011}. It is expected that the addition of heat fluxes will result in flattening of the parallel and perpendicular  temperature profiles along  the entire magnetic nozzle, thus resulting in higher  Mach number at the exit point. These effects require further studies. It would be interesting to implement kinetic closures \cite{SnyderPoP1997} for the heat flux as proposed in Refs. \onlinecite{GuoPoP2014}. The kinetic ion model would also be required to describe the possible ion trapping and demagnetization effects for very low values of the  magnetic field  in the expander region.
We have assumed fully magnetized ions and did not consider the detachment problem which may not be critical for fusion application, but would need to be analyzed for space propulsion applications \cite{BreizmanPoP2008}. 

 It is expected that electron cooling and trapped electrons in the region of the diverging magnetic field (expander) are important \cite{MerinoPSST2018,WethertonPoP2021,SkovorodinPPR2019}. Here, we have focused on ion dynamics and for simplicity assumed isothermal Boltzmann electrons. Effects of general polytropic equation of state for electrons with $\gamma \ne 1$ can be included similarly as in Ref. \onlinecite{SmolyakovPoP2021}. 

  The obtained global solutions do not allow for arbitrary values of plasma velocity at the entrance point of the mirror region and also fully determine the  plasma velocity at the mirror exit (fixed by the mirror ratio). The uniqueness of the  accelerating solutions will therefore  provide  the constraints on the matter and energy outflow through the magnetic nozzle (mirror). These effects were not previously considered either in the context of open mirror or propulsion applications.  These results raise interesting questions of how the plasma solutions in  the source region can be matched to the mirror.  Such a  solution will require the full analysis  of the matter and energy balance, in particular, electron and energy fluxes. Neither of fluid model approximations used here, such as isothermal electrons  with infinite electron heat conductivity, nor the ion CGL model which neglects the ion heat fluxes are fully  adequate for this purpose. Collisionless regimes of interest for most applications also require kinetic analysis. The adjustment (matching) of plasma source with the magnetic nozzle constraints will occur via partial particle reflections and particle  mixing in the source (that have to be described kinetically), possibly enhanced, in part, by non-stationary fluctuations that will occur in regimes when stationary fluid solutions do not exist. It is expected that in general the magnetic nozzle constraints discussed in this paper will reduce plasma losses through the mirror compared to the standard  estimates based on the collisional transitions between trapped and passing (loss cone) regions \cite{MirnovNF1972}. The particle reflections and mixing will likely smooth out the non-monotonous features in the potential that exist in fluid models in some regimes thus increasing the plasma density in the confinement region. The kinetic analysis of such effects is outside of the scope of the present paper and is left for future studies.
  It is also important to note that additional drag due to interactions with  neutrals (such as charge-exchange interactions) can increase the total thrust and thus be beneficial for the electric propulsion applications as shown experimentally and theoretically \cite{MakrinichPoP2009,FruchtmanIEEE2011}.

\begin{acknowledgements}

This work was supported in part by NSERC Canada and the U.S. Air Force Office of Scientific Research FA9550-15-1-0226 and FA9550-21-1-0031. Computational resources were provided by Compute Canada. The authors thank the investors of TAE Technologies and the entire TAE Team for their support.

\end{acknowledgements}

 \section*{Data availability} Data generated in this study is available from the authors upon reasonable request.

\bibliographystyle{unsrt}
\bibliography{References.bib}

\begin{thebibliography}{10}

\bibitem{KaganovichPoP2020}
Igor~D. Kaganovich, Andrei Smolyakov, Yevgeny Raitses, Eduardo Ahedo,
  Ioannis~G. Mikellides, Benjamin Jorns, Francesco Taccogna, Renaud Gueroult,
  Sedina Tsikata, Anne Bourdon, Jean-Pierre Boeuf, Michael Keidar,
  Andrew~Tasman Powis, Mario Merino, Mark Cappelli, Kentaro Hara, Johan~A.
  Carlsson, Nathaniel~J. Fisch, Pascal Chabert, Irina Schweigert, Trevor
  Lafleur, Konstantin Matyash, Alexander~V. Khrabrov, Rod~W. Boswell, and Amnon
  Fruchtman.
\newblock Physics of {E}$\times ${B} discharges relevant to plasma propulsion
  and similar technologies.
\newblock {\em Physics of Plasmas}, 27(12):120601, 2020.

\bibitem{LongmierPSST2011}
B.~W. Longmier, E.~A. Bering, M.~D. Carter, L.~D. Cassady, W.~J. Chancery,
  F.~R.~C. Diaz, T.~W. Glover, N.~Hershkowitz, A.~V. Ilin, G.~E. McCaskill,
  C.~S. Olsen, and J.~P. Squire.
\newblock Ambipolar ion acceleration in an expanding magnetic nozzle.
\newblock {\em Plasma Sources Science \& Technology}, 20(1):015007, 2011.

\bibitem{onofri2017magnetohydrodynamic}
M~Onofri, P~Yushmanov, S~Dettrick, D~Barnes, K~Hubbard, and T~Tajima.
\newblock Magnetohydrodynamic transport characterization of a field reversed
  configuration.
\newblock {\em Physics of Plasmas}, 24(9):092518, 2017.

\bibitem{BufferandPPCF2014}
H.~Bufferand, G.~Ciraolo, G.~Dif-Pradalier, P.~Ghendrih, P.~Tamain,
  Y.~Marandet, and E.~Serre.
\newblock Magnetic geometry and particle source drive of supersonic divertor
  regimes.
\newblock {\em Plasma Physics and Controlled Fusion}, 56(12):122001, 2014.

\bibitem{KirkPPCF2003}
A.~Kirk, W.~Fundamenski, J.~W. Ahn, and G.~Counsell.
\newblock Parallel sol transport in mast and jet: the impact of the mirror
  force.
\newblock {\em Plasma Physics and Controlled Fusion}, 45(8):1445--1463, 2003.

\bibitem{GhendrihPPCF2011}
P.~Ghendrih, K.~Bodi, H.~Bufferand, G.~Chiavassa, G.~Ciraolo, N.~Fedorczak,
  L.~Isoardi, A.~Paredes, Y.~Sarazin, E.~Serre, F.~Schwander, and P.~Tamain.
\newblock Transition to supersonic flows in the edge plasma.
\newblock {\em Plasma Physics and Controlled Fusion}, 53(5):054019, 2011.

\bibitem{morozov1980steady}
AI~Morozov and LS~Solov’ev.
\newblock Steady-state plasma flow in a magnetic field.
\newblock pages 1--103. Springer, 1980.

\bibitem{HooperAIAA1993}
E.~B. Hooper.
\newblock Plasma detachment from a magnetic nozzle.
\newblock {\em Journal of Propulsion and Power}, 9(5):757--763, 1993.

\bibitem{AhedoPoP2010}
E.~Ahedo and M.~Merino.
\newblock Two-dimensional supersonic plasma acceleration in a magnetic nozzle.
\newblock {\em Physics of Plasmas}, 17(7):073501, 2010.

\bibitem{MerinoPoP2016}
M.~Merino and E.~Ahedo.
\newblock Fully magnetized plasma flow in a magnetic nozzle.
\newblock {\em Physics of Plasmas}, 23(2):023506, 2016.

\bibitem{InutakePST2004}
M.~Inutake, K.~Yoshino, S.~Fujimura, H.~Tobari, T.~Yagai, Y.~Hosokawa, R.~Sato,
  K.~Hattori, and A.~Ando.
\newblock Production of a high-mach-number plasma flow for an advanced plasma
  space thruster.
\newblock {\em Plasma Science \& Technology}, 6(6):2541--2545, 2004.

\bibitem{InutakePPCF2007}
M.~Inutake, A.~Ando, K.~Hattori, H.~Tobari, T.~Makita, M.~Shibata,
  Y.~Kasashima, and T.~Komagome.
\newblock Generation of supersonic plasma flows using an applied-field mpd
  arcjet and icrf heating.
\newblock {\em Plasma Physics and Controlled Fusion}, 49(5A):A121--A134, 2007.

\bibitem{GoswamiPoP2014}
Rajiv Goswami, Jean-François Artaud, Frédéric Imbeaux, and Predhiman Kaw.
\newblock Numerical study of transition to supersonic flows in the edge plasma.
\newblock {\em Physics of Plasmas}, 21(7):072510, 2014.

\bibitem{TogoCPP2018}
S.~Togo, D.~Reiser, P.~Borner, M.~Sakamoto, N.~Ezumi, and Y.~Nakashima.
\newblock Benchmarking of b2 code with a one-dimensional plasma fluid model
  incorporating anisotropic ion pressures on simple mirror configurations.
\newblock {\em Plasma and Fusion Research}, 13:3403022, 2018.

\bibitem{TogoNF2019}
S.~Togo, T.~Takizuka, D.~Reiser, M.~Sakamoto, N.~Ezumi, Y.~Ogawa, K.~Nojiri,
  K.~Ibano, Y.~Li, and Y.~Nakashima.
\newblock Self-consistent simulation of supersonic plasma flows in advanced
  divertors.
\newblock {\em Nuclear Fusion}, 59(7):076041, 2019.

\bibitem{SmolyakovPoP2021}
A.~I. Smolyakov, A.~Sabo, P.~Yushmanov, and S.~Putvinskii.
\newblock On quasineutral plasma flow in the magnetic nozzle.
\newblock {\em Physics of Plasmas}, 28(6):060701, 2021.

\bibitem{chew1956boltzmann}
GF~Chew, ML~Goldberger, and FE~Low.
\newblock The boltzmann equation and the one-fluid hydromagnetic equations in
  the absence of particle collisions.
\newblock {\em Proceedings of the Royal Society of London. Series A.
  Mathematical and Physical Sciences}, 236(1204):112--118, 1956.

\bibitem{zawaideh1986generalized}
Emad Zawaideh, Farrokh Najmabadi, and Robert~W Conn.
\newblock Generalized fluid equations for parallel transport in collisional to
  weakly collisional plasmas.
\newblock {\em The Physics of fluids}, 29(2):463--474, 1986.

\bibitem{GuoPoP2014}
Zehua Guo, Xian-Zhu Tang, and Chris McDevitt.
\newblock Parallel heat flux and flow acceleration in open field line plasmas
  with magnetic trapping.
\newblock {\em Physics of Plasmas}, 21(10):102512, 2014.

\bibitem{SmolyakovPoP2010}
A.~I. Smolyakov and X.~Garbet.
\newblock Drift kinetic equation in the moving reference frame and reduced
  magnetohydrodynamic equations.
\newblock {\em Physics of Plasmas}, 17(4):042105, 2010.

\bibitem{SnyderPoP1997}
P.~B. Snyder, G.~W. Hammett, and W.~Dorland.
\newblock Landau fluid models of collisionless magnetohydrodynamics.
\newblock {\em Physics of Plasmas}, 4(11):3974--3985, 1997.

\bibitem{RobertsonPoP2016}
S.~Robertson.
\newblock A reduced set of gyrofluid equations for plasma flow in a diverging
  magnetic field.
\newblock {\em Physics of Plasmas}, 23(4):043513, 2016.

\bibitem{HelanderPoP1994}
P.~Helander, S.~I. Krasheninnikov, and P.~J. Catto.
\newblock Fluid equations for a partially-ionized plasma.
\newblock {\em Physics of Plasmas}, 1(10):3174--3180, 1994.

\bibitem{NgPoP2007}
Sheung-Wah Ng and A.~B. Hassam.
\newblock Neutral penetration in centrifugally confined plasmas.
\newblock {\em Physics of Plasmas}, 14(10):102508, 2007.

\bibitem{LehnertNF1971}
B.~Lehnert.
\newblock Rotating plasmas.
\newblock {\em Nuclear Fusion}, 11(5):485--533, 1971.

\bibitem{TogoNME2019}
S.~Togo, T.~Takizuka, D.~Reiser, M.~Sakamoto, Y.~Ogawa, N.~Ezumi, K.~Ibano,
  K.~Nojiri, Y.~Li, and Y.~Nakashima.
\newblock Characteristics of plasma flow profiles in a super-x-divertor-like
  configuration.
\newblock {\em Nuclear Materials and Energy}, 19:149--154, 2019.

\bibitem{LittlePRL2016}
J.~M. Little and E.~Y. Choueiri.
\newblock Electron cooling in a magnetically expanding plasma.
\newblock {\em Physical Review Letters}, 117(22):225003, 2016.

\bibitem{MerinoPSST2020}
Mario Merino, Pablo Fajardo, Gabriel Giono, Nickolay Ivchenko, Jon-Tomas
  Gudmundsson, Stephane Mazouffre, Dimitry Loubere, and Kathe Dannenmayer.
\newblock Collisionless electron cooling in a plasma thruster plume:
  experimental validation of a kinetic model.
\newblock {\em Plasma Sources Science \& Technology}, 29(3):035029, 2020.

\bibitem{MirnovNF1972}
V.~V. Mirnov and D.~D. Ryutov.
\newblock Gas-dynamic description of a plasma in a corrugated magnetic-field.
\newblock {\em Nuclear Fusion}, 12(6):627--636, 1972.

\bibitem{SkovorodinPPR2012}
D.~I. Skovorodin and A.~D. Beklemishev.
\newblock Plasma outflow from a corrugated trap in the kinetic regime.
\newblock {\em Plasma Physics Reports}, 38(3):202--206, 2012.

\bibitem{PekkerSovJPP1984}
M.~S. Pekker.
\newblock Particle scattering by potential jumps in a tandem mirror with a
  thermal barrier.
\newblock {\em Sov. J. Plasma Physics}, 10(11):33--35, 1984.

\bibitem{ManheimerIEEE2001}
W.~M. Manheimer and R.~F. Fernsler.
\newblock Plasma acceleration by area expansion.
\newblock {\em Ieee Transactions on Plasma Science}, 29(1):75--84, 2001.

\bibitem{FruchtmanPoP2012}
A.~Fruchtman, K.~Takahashi, C.~Charles, and R.~W. Boswell.
\newblock A magnetic nozzle calculation of the force on a plasma.
\newblock {\em Physics of Plasmas}, 19(3):033507, 2012.

\bibitem{ParkerAJ1958}
E.~N. Parker.
\newblock Dynamics of the interplanetary gas and magnetic fields.
\newblock {\em Astrophysical Journal}, 128(3):664--676, 1958.

\bibitem{DubinovJPP2005}
Alexander~E. Dubinov and Irina~D. Dubinova.
\newblock How can one solve exactly some problems in plasma theory.
\newblock {\em Journal of Plasma Physics}, 71(05):715, 2005.

\bibitem{RaimbaultPoP2007}
J.~L. Raimbault, L.~Liard, J.~M. Rax, P.~Chabert, A.~Fruchtman, and
  G.~Makrinich.
\newblock Steady-state isothermal bounded plasma with neutral dynamics.
\newblock {\em Physics of Plasmas}, 14(1):013503, 2007.

\bibitem{DubinovPoP2022}
Alexander~E. Dubinov.
\newblock Mathematical tricks for pseudopotentials in the theories of nonlinear
  waves in plasma.
\newblock {\em Physics of Plasmas}, 2022.

\bibitem{CranmerAJP2004}
Steven~R. Cranmer.
\newblock New views of the solar wind with the lambert w function.
\newblock {\em American Journal of Physics}, 72(11):1397--1403, 2004.

\bibitem{BinderbauerAIP2016}
M.~W. Binderbauer, T.~Tajima, M.~Tuszewski, L.~Schmitz, A.~Smirnov, H.~Gota,
  E.~Garate, D.~Barnes, B.~H. Deng, E.~Trask, X.~Yang, S.~Putvinski, R.~Andow,
  N.~Bolte, D.~Q. Bui, F.~Ceccherini, R.~Clary, A.~H. Cheung, K.~D. Conroy,
  S.~A. Dettrick, J.~D. Douglass, P.~Feng, L.~Galeotti, F.~Giammanco,
  E.~Granstedt, D.~Gupta, S.~Gupta, A.~A. Ivanov, J.~S. Kinley, K.~Knapp,
  S.~Korepanov, M.~Hollins, R.~Magee, R.~Mendoza, Y.~Mok, A.~Necas,
  S.~Primavera, M.~Onofri, D.~Osin, N.~Rath, T.~Roche, J.~Romero, J.~H.
  Schroeder, L.~Sevier, A.~Sibley, Y.~Song, L.~C. Steinhauer, M.~C. Thompson,
  A.~D. Van~Drie, J.~K. Walters, W.~Waggoner, P.~Yushmanov, K.~Zhai, and
  T.~A.~E. Team.
\newblock Recent breakthroughs on c-2u: Norman's legacy.
\newblock In T.~Tajima and M.~Binderbauer, editors, {\em Physics of
  Plasma-Driven Accelerators and Accelerator-Driven Fusion}, volume 1721 of
  {\em AIP Conference Proceedings}, page 030003, 2016.

\bibitem{MakrinichPoP2009}
G.~Makrinich and A.~Fruchtman.
\newblock Experimental study of a radial plasma source.
\newblock {\em Physics of Plasmas}, 16(4):043507, 2009.

\bibitem{FruchtmanJPhysD2017}
A.~Fruchtman.
\newblock Neutral gas depletion in low temperature plasma.
\newblock {\em Journal of Physics D-Applied Physics}, 50(47):473002, 2017.

\bibitem{FruchtmanIEEE2011}
A.~Fruchtman.
\newblock The thrust of a collisional-plasma source.
\newblock {\em Ieee Transactions on Plasma Science}, 39(1):530--539, 2011.

\bibitem{lieberman2005principles}
M.~Lieberman and A.~Lichtenberg.
\newblock {\em Principles of plasma discharges and materials processing}.
\newblock (Wiley-Blackwell, 2005).

\bibitem{janev1987collision}
Ratko~K Janev, William~D Langer, Douglass~E Post, and Kenneth Evans.
\newblock {\em Elementary Processes in Hydrogen-Helium Plasmas}.
\newblock (Springer, 1987).

\bibitem{BreizmanPoP2008}
B.~N. Breizman, M.~R. Tushentsov, and A.~V. Arefiev.
\newblock Magnetic nozzle and plasma detachment model for a steady-state flow.
\newblock {\em Physics of Plasmas}, 15(5):057103, 2008.

\bibitem{WachsPSST2020}
B.~Wachs and B.~Jorns.
\newblock Background pressure effects on ion dynamics in a low-power magnetic
  nozzle thruster.
\newblock {\em Plasma Sources Science \& Technology}, 29(4):045002, 2020.

\bibitem{TakahashiAPL2016}
K.~Takahashi, Y.~Takao, and A.~Ando.
\newblock Neutral-depletion-induced axially asymmetric density in a helicon
  source and imparted thrust.
\newblock {\em Applied Physics Letters}, 108(7):074103, 2016.

\bibitem{MerinoPSST2018}
M.~Merino, J.~Maurino, and E.~Ahedo.
\newblock Kinetic electron model for plasma thruster plumes.
\newblock {\em Plasma Sources Science and Technology}, 27(3):035013, 2018.

\bibitem{AhedoPSST2020}
E.~Ahedo, S.~Correyero, J.~Navarro-Cavallé, and M.~Merino.
\newblock Macroscopic and parametric study of a kinetic plasma expansion in a
  paraxial magnetic nozzle.
\newblock {\em Plasma Sources Science and Technology}, 29(4):045017, 2020.

\bibitem{MartinezPoP2011}
Manuel Martinez-Sanchez and Eduardo Ahedo.
\newblock Magnetic mirror effects on a collisionless plasma in a convergent
  geometry.
\newblock {\em Physics of Plasmas}, 18(3):033509, 2011.

\bibitem{WethertonPoP2021}
B.~A. Wetherton, A.~Le, J.~Egedal, C.~Forest, W.~Daughton, A.~Stanier, and
  S.~Boldyrev.
\newblock A drift kinetic model for the expander region of a magnetic mirror.
\newblock {\em Physics of Plasmas}, 28(4):042510, 2021.

\bibitem{SkovorodinPPR2019}
D.~I. Skovorodin.
\newblock Influence of trapped electrons on the plasma potential in the
  expander of an open trap.
\newblock {\em Plasma Physics Reports}, 45(9):799--804, 2019.

\end{thebibliography}

\appendix
\section{Magnetic field profile}

To study the effects of finite ion temperature we employed a mirror magnetic field. To simplify numerical calculations, the magnetic field was described by three different
functions in regions A, B, and C: the region A, from $z^{\prime }=z/L=0$ to $%
z^{^{\prime }}=$0.33, with $B_{A}(z')=0.5e^{-84(z^{\prime}-0.5)^2}+B_{0}$, $%
B_{0}=0.067$ (T); the region B from $z^{^{\prime }}=$0.33 to $z^{^{\prime
}}=$0.77 with $B_{B}(z')=B_{m}\left(0.13\right)^{3}/\left(\left(0.13\right)^{2}+(z^{\prime}-z_{m})^{2})^{\frac{3}{2}}\right),$
where $B_{m}=0.5365 (T)$ is the magnetic field at the maximum at $z^{\prime}=z_{m}$; the regions C from $z^{^{\prime }}=$0.77 to $z^{^{\prime }}=$1 with $%
B_{C}(z')=0.55e^{-51(z^{\prime}-0.5)}+B_{l},$ $B_{l}=0.0268$ (T). The functions $%
B_{A}(z')$, $B_{B}(z')$, and $B_{C}(z')$ are chosen to have $B$ and
\thinspace $dB/dz$ continuous across the boundaries $A-B$ and $B-C$.

\section{Time-dependent equations}

The solutions of the stationary equations were verified with the initial value simulations of the time-dependent equations. In the absence of charge-exchange and ionization they have the following form:
\begin{equation}
\begin{split}
\frac{\partial n}{\partial t}= nM \frac{\partial \ln B}{\partial
z}- M\frac{\partial n}{\partial z}-n\frac{\partial M }{\partial z} + \alpha_{1}\frac{\partial^2 n}{\partial z^2},   
\end{split}
\label{n-BOUT}
\end{equation}

\begin{equation}
\begin{split}
\frac{\partial p_{\parallel }}{\partial t}=  p_{\parallel }M \frac{%
\partial \ln B}{\partial z}- M\frac{\partial p_\parallel} {%
\partial z} -3p_{\parallel }\frac{\partial M}{%
\partial z}  + \alpha_{2}\frac{\partial^2 p_{\parallel}}{\partial z^2},
\end{split}
\label{p-para-BOUT}
\end{equation}

\begin{equation}
\begin{split}
\frac{\partial p_{\perp }}{\partial t}=  2p_{\perp }M\frac{%
\partial \ln B}{\partial z}-M\frac{\partial p_{\perp }}{
\partial z}-p_{\perp }\frac{\partial M}{\partial z} 
 + \alpha_{3}\frac{\partial^2 p_{\perp}}{\partial z^2},
\end{split}
\label{p-perp-BOUT}
\end{equation}

\begin{equation}
\begin{split}
\frac{\partial M}{\partial t}= & -M\frac{%
\partial M}{\partial z} - \frac{1}{n}\frac{\partial n}{\partial z} -\frac{1}{n}\left(\frac{T_{i\parallel0}}{T_e}\right)\frac{\partial p_{\parallel }}{\partial z}+ \\
& \frac{1}{n}\left(\left(\frac{T_{i\parallel0}}{T_e}\right) p_{\parallel
} - \left(\frac{T_{i\perp0}}{T_e}\right) p_{\perp}\right) \frac{\partial \ln B}{\partial z}
+ \alpha_{4}\frac{\partial^2 M}{\partial z^2}.
\end{split}
\label{M-BOUT}
\end{equation}
Here, with the exception of $T_{i\parallel0}$, $T_{i\perp0}$ and $T_e$, all quantities are  expressed in dimensionless units. The diffusive coefficients $\alpha$ had the following values: \
$\alpha_1 = 5.0 \times 10^{-6}$, $\alpha_2 = 5.0 \times 10^{-6}$, $\alpha_3 = 5.0 \times 10^{-6}$ and $\alpha_4 = 1.0 \times 10^{-9}$.

Note that while the profiles of the  density and pressure are fully determined by the global solution, the absolute values have free normalization parameters. Plasma parameters are normalized to their respective values
at the left end of the nozzle such that $n^{\prime }=n/n_{0}$, $%
p_{\parallel }^{\prime }=p_{\parallel }/p_{\parallel_0}$, $%
p_{\perp }^{\prime }=p_{\perp }/p_{\perp_0}$, $T^{\prime
}_{i_\parallel }=T_{i_\parallel }/T_{e}$, $T^{\prime
}_{i_\perp }=T_{i_\perp }/T_{e}$, $z^{\prime }=z/L$ and $%
t^{\prime }=c_{s}t/L$. (For the
sake of convenience all the primes on parameters will be dropped and it will
be assumed that n, $p_{\parallel }$, $p_{\perp }$, $\phi $, $%
T_{i}{}_{\parallel }$, $T_{i}{}_{\perp }$, $z$ and $t$ represent
normalized quantities).

The values for the diffusion coefficients $\alpha$ were small in comparison to the other terms in the equation and thus their addition did not affect the physics of the problem. For instance, for a value of $\alpha= 5.0 \times 10^{-6}$, $M=3.179 \times 10^{-4}$ and $L=4$ the condition $M/L = 7.95 \times 10^{-5} >>\alpha/L^2 = 3.13 \times 10^{-7}$ holds.    

When ionization and charge-exchange effects were included in the model, the time-dependent equations  had the form
\begin{equation}
\begin{split}
\frac{\partial n}{\partial t}= & nM \frac{\partial \ln B}{\partial
z}- M\frac{\partial n}{\partial z}-n\frac{\partial M }{\partial z} + \nu_{1} n \\
& + \beta_{1}\frac{\partial^2 n}{\partial z^2},   
\end{split}
\label{n-cx-BOUT}
\end{equation}

\begin{equation}
\begin{split}
\frac{\partial p_{\parallel}}{\partial t}= & p_{\parallel}M \frac{%
\partial \ln B}{\partial z}- M\frac{\partial p_\parallel} {%
\partial z} -3p_{\parallel}\frac{\partial M}{%
\partial z} - \nu_{2} p_{\parallel}\\
& + \beta_{2}\frac{\partial^2 p_{\parallel}}{\partial z^2},
\end{split}
\label{p-para-cx-BOUT}
\end{equation}

\begin{equation}
\begin{split}
\frac{\partial p_{\perp}}{\partial t}= & 2p_{\perp }M\frac{%
\partial \ln B}{\partial z} - M\frac{\partial p_{\perp}}{
\partial z}-p_{\perp }\frac{\partial M}{\partial z} - \nu_{2} p_{\perp}\\
& + \beta_{3}\frac{\partial^2 p_{\perp}}{\partial z^2},
\end{split}
\label{p-perp-cx-BOUT}
\end{equation}

\begin{equation}
\begin{split}
\frac{\partial M}{\partial t}= & -M\frac{%
\partial M}{\partial z} - \frac{1}{n}\frac{\partial n}{\partial z} -\frac{1}{n}\left(\frac{T_{i\parallel0}}{T_e}\right)\frac{\partial p_{\parallel }}{\partial z}+ \\
& \frac{1}{n}\left(\left(\frac{T_{i\parallel0}}{T_e}\right) p_{\parallel
} - \left(\frac{T_{i\perp0}}{T_e}\right) p_{\perp }\right) \frac{\partial \ln B}{\partial z} - \nu_{2} M + \beta_{4}\frac{\partial^2 M}{\partial z^2}.
\end{split}
\label{M-cx-BOUT}
\end{equation}

The diffusion coefficients $\beta$ were dimensionless and had the following values: $\beta_1 = 1.0 \times 10^{-3}$,  $\beta_2 = 6.0 \times 10^{-3}$,  $\beta_3 = 9.9 \times 10^{-4}$ and  $\beta_4 = 1.0 \times 10^{-4}$.

Similar to the values of $\alpha$, the values of $\beta$ were small and did not affect the physics of the problem. For instance for $\beta=1.0 \times 10^{-3}$, $M=0.127$ and $L=4$ the condition $M/L = 0.032 >> \beta/L^2 = 6.25 \times 10^{-5}$ is true and the addition of the parameters $\beta$ did not affect the physics of the problem.

The time-dependent equations describing the flow of plasma were solved in BOUT++. In the BOUT++ simulations, $p_\parallel$ was displaying
oscillatory behavior in region A of the nozzle. These oscillations were
damped as $t$ increased and the time-dependent value of $p_\parallel$
approached the stationary solution value.


\end{document}